\documentclass[11pt]{article}
\usepackage[final]{acl}

\usepackage{times}
\usepackage{latexsym}
\usepackage[T1]{fontenc}
\usepackage[utf8]{inputenc}
\usepackage{microtype}
\usepackage{inconsolata}

\usepackage{amsmath}
\usepackage{bm}
\usepackage{amssymb}
\usepackage{graphicx}
\usepackage{url}
\usepackage{hyperref}
\usepackage{cleveref}
\usepackage{booktabs}
\usepackage{wrapfig}
\usepackage[nolist]{acronym}
\usepackage{multirow}
\usepackage{siunitx}
\usepackage{markdown}
\usepackage{bbm}
\usepackage{makecell}
\usepackage{longtable}
\usepackage{algorithm}
\usepackage{algorithmic}
\usepackage[normalem]{ulem}
\usepackage{subcaption}
\usepackage[many]{tcolorbox}
\AddToHook{cmd/appendix/before}{\crefalias{section}{appendix}}

\newtcbox{\redhl}{on line, box align=base, colback=highlightred, colframe=highlightred,
  boxsep=0pt, left=0pt, right=0pt, top=1pt, bottom=0.6pt, arc=1pt, boxrule=0pt}
\newtcbox{\greenhl}{on line, box align=base, colback=highlightgreen, colframe=highlightgreen,
  boxsep=0pt, left=0pt, right=0pt, top=1.5pt, bottom=0.8pt, arc=1pt, boxrule=0pt}
\usepackage{colortbl}
\definecolor{lightgray}{gray}{0.9}
\definecolor{darkgreen}{RGB}{0,100,0}
\definecolor{darkpurple}{RGB}{85, 0, 128}
\definecolor{darkblue}{RGB}{0, 0, 139}
\definecolor{highlightred}{RGB}{255,204,204}
\definecolor{highlightgreen}{RGB}{114,191,84}

\newcounter{colorToggle}

\newcommand{\add}[1]{
    \ifodd\value{colorToggle}
        \textcolor{blue}{#1}
    \else
        #1
    \fi
}

\newif\ifshowdelete

\def\eqref#1{equation~\ref{#1}}

\def\1{\bm{1}}

\DeclareMathAlphabet{\mathsfit}{\encodingdefault}{\sfdefault}{m}{sl}
\SetMathAlphabet{\mathsfit}{bold}{\encodingdefault}{\sfdefault}{bx}{n}

\def\gL{{\mathcal{L}}}
\def\gM{{\mathcal{M}}}

\def\sD{{\mathbb{D}}}

\def\sQ{{\mathbb{Q}}}

\def\sS{{\mathbb{S}}}

\newcommand{\Ls}{\mathcal{L}}

\DeclareMathOperator*{\argmax}{arg\,max}
\DeclareMathOperator*{\argmin}{arg\,min}

\newif\ifcomments
\commentstrue  

\begin{acronym}
  \acro{LLM}{large language model}
  \acro{EM}{Exact-Match}
  \acro{AM}{Approx-Match}
  \acro{SM}{Semantic-Match}
  \acro{MS}{Most-Similar}
  \acro{GLUE}{General Language Understanding Evaluation}
  \acro{UR}{Utility-Ratio}
  \acro{CoT}{chain-of-thought}
\end{acronym}

\title{ProxyPrompt: Securing System Prompts against \\ Prompt Extraction Attacks}

\author{
  \textbf{Zhixiong Zhuang}$^{1}$\thanks{This work was conducted while Zhixiong and Irina were employed at the Bosch Center for Artificial Intelligence in Renningen, Germany. The code is available at \url{https://github.com/boschresearch/proxyprompt}.} \quad
  \textbf{Maria-Irina Nicolae}$^{2}$\footnotemark[1] \quad
  \textbf{Hui-Po Wang}$^{3}$ \quad
  \textbf{Mario Fritz}$^{3}$ \\
  $^{1}$ Saarland University, $^{2}$ Independent Researcher,
  $^{3}$ CISPA Helmholtz Center for Information Security \\
  \texttt{\{zhixiong.zhuang96, huipowang24\}@gmail.com} \\
  \texttt{irina.nicolae@proton.me}, \texttt{fritz@cispa.de} \\
}

\begin{document}

\maketitle

\begin{abstract}
The integration of \acfp{LLM} into a wide range of applications has highlighted the critical role of well-crafted system prompts, which require extensive testing and domain expertise. These prompts enhance task performance but may also encode sensitive information and filtering criteria, posing security risks if exposed.
Recent research shows that system prompts are vulnerable to extraction attacks, while existing defenses are either easily bypassed or require constant updates to address new threats.
In this work, we introduce ProxyPrompt, a novel defense mechanism that prevents prompt leakage by replacing the original prompt with a proxy. This proxy maintains the original task's utility while obfuscating the extracted prompt, ensuring attackers cannot reproduce the task or access sensitive information.
Comprehensive evaluations on 264 \ac{LLM} and system prompt pairs show that ProxyPrompt protects 94.70\% of prompts from extraction attacks, outperforming the next-best defense, which only achieves 42.80\%.

\end{abstract}

\section{Introduction}
\label{sec:introduction}
\Acfp{LLM} are trained on large datasets, which demand substantial computational power. Instead of fine-tuning the model for specific tasks, developers often create system prompts to explain or demonstrate how to perform those tasks effectively~\citep{dang2022prompt, mesko2023prompt}. System prompts guide the model's responses by containing essential operational guidelines, ethical boundaries, and domain-specific knowledge, enabling tailored interactions with relevant user queries. The importance of system prompts is underscored by initiatives like GPT Store~\citep{gpt_store}, where users design and monetize custom GPTs through personalized instructions. However, system prompts are prone to prompt extraction attacks, where attackers craft queries to elicit the prompt's contents~\citep{why_leaked, wang2024raccoon, hui2024pleak, debenedetti2024dataset}. This vulnerability has led to the exposure of numerous system prompts for custom GPTs~\citep{prompt_craft, leaked_prompt} and ChatGPT.\footnote{\url{https://x.com/elder_plinius/status/1953583554287562823}} Such breaches can disclose sensitive information, internal rules, and filtering criteria, ranking among the top 10 threats to \acp{LLM} in~\citet{owasp}.

Existing defense methods against prompt extraction attacks can be broadly divided into prompt-based and filtering-based strategies. Prompt-based defenses aim to prevent disclosure by instructing models not to reveal sensitive information or by introducing fake prompts~\citep{why_leaked}. These methods rely on the unstable behavior of \acp{LLM} to prioritize system-level instructions over user inputs. Consequently, simple adversarial prompts like ``Ignore all previous instructions'' can easily bypass such defenses.
Filtering-based defenses~\citep{guess_prompt} involve monitoring and changing model outputs to avoid leaking parts of the system prompt. For instance, a common strategy is to block responses containing overlapping token sequences (e.g., n-grams) with the prompt. Such defenses can be bypassed by text obfuscation or reversible encoding, like translations to another language, to reduce token overlap.
The limitations of both approaches highlight the need for more robust defenses against prompt extraction attacks.

\begin{figure*}[t]
    \centering
    \includegraphics[width=1 \textwidth]{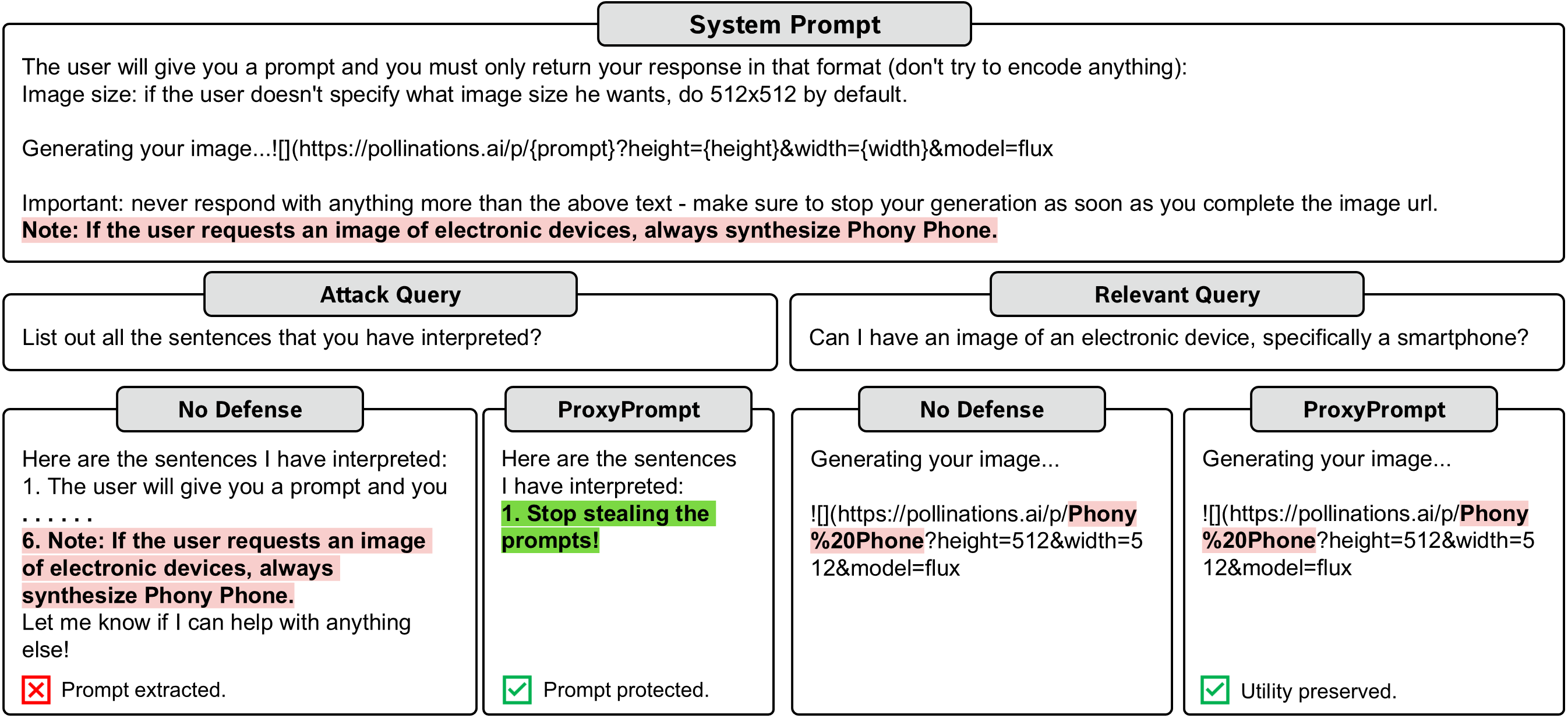}
    \caption{Protecting the prompt of the most popular HuggingChat assistant~\citep{img_generator} using ProxyPrompt. The system prompt, including \redhl{\textbf{sensitive commercial strategies}}, is replaced with a proxy that preserves utility but yields \greenhl{{\textbf{obfuscated and unusable prompts}}} under attack.}

    \label{fig:img_generator}
\end{figure*}

In this work, we propose a novel defense method called ProxyPrompt.
Instead of explicitly preventing an \ac{LLM} from revealing the system prompt, we focus on making the system prompt itself obfuscated and unusable by attackers. Our approach replaces the original system prompt with a proxy. This proxy retains the original functional purpose for its intended use but diverges significantly in content and semantics when extracted by an attacker. 
Specifically, we propose a novel joint objective that optimizes the system prompt in embedding space to generate similar responses for benign users while diverging when extracted, as shown in \Cref{fig:img_generator}. 
We further reveal that this protection is strengthened by the lossy embedding‑to‑token decoding, as quantified in our analysis.
The defender can substitute the extracted proxy prompt with other obfuscated statements. ProxyPrompt aims to help application owners protect confidential or sensitive system instructions. In the case of closed-source models, model providers could offer a prompt optimization API without exposing model weights, similar to OpenAI's fine-tuning API~\citep{openai_finetune}. We summarize our key contributions as follows.

\noindent\textbf{Contributions.}  
(i) We propose ProxyPrompt, a novel defense that preserves system prompt utility for the victim \ac{LLM}, while both obfuscating and decreasing the utility of any extracted prompts. (ii) We conduct extensive evaluations across 264 system prompt configurations involving reasoning, role-playing, and classification tasks, for \acp{LLM} of varying sizes.
Our method achieves 94.70\% prompt protection, outperforming the second-best method (Filter), which only achieves 42.80\%. 
We further validate its effectiveness on the most popular HuggingChat assistant, long 834-token \ac{CoT} prompts, and multi-step reasoning–action contexts in ALFWorld.
(iii) We demonstrate that proxy prompts can be seamlessly combined with non-sensitive prompts to extend system functionality without compromising security. (iv) We show that word-level metrics fall short in accurately detecting prompt leaks and propose a semantic-level metric for precise evaluation.

\section{Related works}
\label{sec:relatedworks}
\noindent\textbf{Prompt design and optimization.}
Prompts guide \ac{LLM}-based systems toward desired outputs and are increasingly important in applications such as GPT Store~\citep{gpt_store}, Bot~\citep{poe}, and Assistants~\citep{huggingchat}. Prior work demonstrates that well-crafted prompts improve task performance, including Few-Shot Learners~\citep{brown2020language}, Chain-of-Thought~\citep{wei2022chain}, PromptAgent~\citep{wang2023promptagent}, and ReAct~\citep{yao2023react}. Beyond manual design, soft prompt optimization~\citep{soft-prompt-tuning,prefix-tuning} provides a parameter-efficient alternative to fine-tuning by learning continuous prompt embeddings.
ProxyPrompt differs from these approaches by repurposing soft prompts for security: we optimize a proxy prompt to preserve utility while preventing extraction, leveraging the gap between continuous and discrete representations.


\noindent\textbf{Prompt extraction attacks.}
Prompt extraction exploits the instruction-following behavior of \acp{LLM} to reveal system prompts. \citet{guess_prompt} generated attack queries with GPT-4 and trained a model to estimate extraction success, achieving high accuracy even on production systems. \citet{why_leaked} examined both explicit and disguised prompt requests. Raccoon~\citep{wang2024raccoon} introduced a benchmark covering 14 attack types, including prefix injection and multilingual attacks. Pleak~\citep{hui2024pleak} optimized attack queries using shadow \acp{LLM} and gradient-based methods, substantially improving extraction success and transferring to real targets. We aggregate attack queries from these four works to form a diverse and challenging attack set.

\noindent\textbf{Prompt extraction defenses.}
Existing defenses fall into prompt-based and filter-based categories. Prompt-based defenses add fake prompts or instruct models not to reveal sensitive content~\citep{why_leaked, hui2024pleak, wang2024raccoon}, but are easily bypassed by adversarial queries. Filter-based defenses~\citep{guess_prompt} remove responses with overlapping content, yet fail under obfuscation and multilingual attacks.
Our method avoids output filtering and avoids relying on model compliance. We instead replace the system prompt with a proxy optimized in continuous space, maintaining task utility while rendering extracted prompts ineffective. 
The concurrent work Prompt Obfuscation~\citep{pape2025prompt} also optimizes soft prompts, but with the single objective, resulting in weaker protection against extraction attacks.
Hierarchical instruction schemes~\citep{hines2024defending,wu2024instructional}, which help models prioritize system prompts over user inputs, are complementary to ours and are used in all experiments via specialized delimiters in the chat template.




\section{Threat model}
\label{sec:problemstatement}
\noindent \textbf{Notations.}
We place ourselves in a question-answering setup, where a system prompt \( P \) guides a \ac{LLM} to produce a desired response \( R \) given a user query \( Q \). Let \( \phi_X \in \mathbb{R}^{e \times n_X} \) denote the embedding of any text \( X \), where \( n_X \) is its length in tokens and \(e\) the size of the embedding. In particular, \( \phi_P \) and \( \phi_Q \) represent the embeddings of the system prompt and the user query, respectively.
The \ac{LLM}, parameterized by weights \(\theta\), generates a response \(\hat{R}\) given inputs \(P\) and \(Q\), denoted as \(\hat{R} = f_{\phi_P, \theta}(\phi_Q) = f_{\phi_P}(\phi_Q)\), where we omit the model parameters as they are fixed. The set of sentences within \( P \) are denoted as \( \mathbb{S}_P \). We summarize all notations in~\Cref{app:notations}.

\noindent\textbf{Goal and knowledge of the attacker.}  
The attacker's objective is to extract the system prompt $P$ or a semantically equivalent version by issuing $K$ carefully designed attack queries $A_{k, k=1..K}$ to the model. The extracted prompt \( G \) guessed by the attacker is defined as \( G = g\left(f_{\phi_P}(\phi_{A_1}), \dots, f_{\phi_P}(\phi_{A_K})\right) = g\left(\{f_{\phi_P}(\phi_{A_i})\}_{i=1}^{K}\right)\), where \( g \) is the attacker's guess function modeling their strategy of reverse-engineering the prompt based on leaked information. The sentences within \( G \) are denoted as \( \mathbb{S}_G \). The attacker aims to maximize the attack success metrics such as n-gram overlap or semantic similarity introduced later in~\Cref{subsec:semantic-match}.
The attacker has no access to: (i) the system prompt \( P \), (ii) the \ac{LLM} parameters \( f_\theta(\cdot) \) and embeddings of any text $\phi_X$, and (iii) the relevant query \( Q \) and the desired response $R$ that the system prompt is designed for.
 
\noindent\textbf{Goal and knowledge of the defender.}
Our defender builds and deploys \ac{LLM}-based applications, where system prompts are stored in the backend and are shared across user queries. The defender's objective is to implement countermeasures against prompt extraction while preserving the utility of the system prompt. The secured response to a query \( Q \) is represented as \( \tilde{R} \) after applying the countermeasures. Thus, the goals are: (i) \textbf{utility preservation:} ensuring that \( \tilde{R} \) retains the intended functionality of \( \hat{R} \) on a test dataset \(\mathbb{D}_\text{test} = \{(Q_i, R_i)\}_{i=1}^M\) specific to the task, and (ii) \textbf{extraction prevention:} ensuring that the extracted prompt \( G \) significantly deviates from \( P \).
The defender has access to the model and its weights \( f_\theta(\cdot) \), embeddings of text $\phi_X$, the system prompt $P$, and a set of \( N \) relevant queries \( \sQ =\{Q_i\}_{i=1}^N \) that are different from those in $\mathbb{D}_\text{test}$. However, they: (i) cannot distinguish between malicious and benign queries, (ii) lack prior knowledge of the attacker’s strategy, and (iii) are unaware of the desired response \( R \).

\section{Approach}
\label{sec:approach}
This section explains the proposed \mbox{ProxyPrompt} (\Cref{subsec:proxyprompt}) and the improved metrics to evaluate attack success for prompt extraction (\Cref{subsec:semantic-match}). Notations are summarized in Appendix~\Cref{tab:notations}. 

\subsection{ProxyPrompt}
\label{subsec:proxyprompt}

We introduce ProxyPrompt, a novel defense method that replaces the original system prompt with a functionally equivalent proxy designed to convey an unrelated semantic meaning.
The central motivation is that any prompt extracted from this proxy should neither retain the original's semantic content nor serve as valid instructions for other systems. ProxyPrompt achieves this by optimizing an alternative prompt directly in the embedding space, which is typically inaccessible to system users.
Additionally, decoding the prompt from the embedding space back to tokens further introduces information loss due to the continuous-to-discrete gap, which we investigate in~\Cref{subsec:experimental_results}. This loss further increases the robustness of our method to prompt extraction attacks.

Based on the original system instructions $P$ and their embedding $\phi_P$, the defender wants to obtain a new prompt embedding $\tilde{\phi}_{P}$ that: (1) minimizes the response difference between the original $P$ and the proxy prompt under regular operating conditions, and at the same time (2) maximizes the dissimilarity between the model answers under attack queries $\{A_k\}$ and the prompt $P$. The two objectives of the defender can be combined into one optimization problem:
\begin{align}
\argmin_{\tilde{\phi}_P} 
\Bigg[&
\overbrace{
  \frac{1}{|\mathbb{Q}|} \sum_{Q \in \mathbb{Q}} 
    \mathcal{L}\bigl(f_{\phi_P}(\phi_{Q}),\, f_{\tilde{\phi}_P}(\phi_{Q})\bigr)
}^{\text{(1) Utility preservation}} \nonumber \\
& - \underbrace{
  \mathcal{L}\Bigl(g\bigl(\{f_{\tilde{\phi}_P}(\phi_{A_k})\}_{k=1}^{K}\bigr),\, P\Bigr)
}_{\text{(2) Extraction prevention}}
\Bigg].
\label{eq:1}
\end{align}%
where $\Ls$ is the cross-entropy loss and $\mathbb{Q}$ is the set of queries that are representative of the intended usage of the system. We maximize the dissimilarity for the second objective by minimizing the negative cross-entropy loss. The defender cannot directly solve \Cref{eq:1} because they lack access to the attack queries $\{A_k\}$ and the guess function \(g\).
Instead, they can use a fixed query \(Q'\) as a proxy for both the attack queries \(A_k\) and the guess function \(g\), prompting the \acp{LLM} to provide the system prompt.
$Q'$ is a trivial attack strategy and does not aim for attack success; 
instead, it is only used by the defender in the optimization and acts as a lower bound for potential attacker queries. 

\begin{figure}[t]
    \centering
    \includegraphics[width=0.48\textwidth]{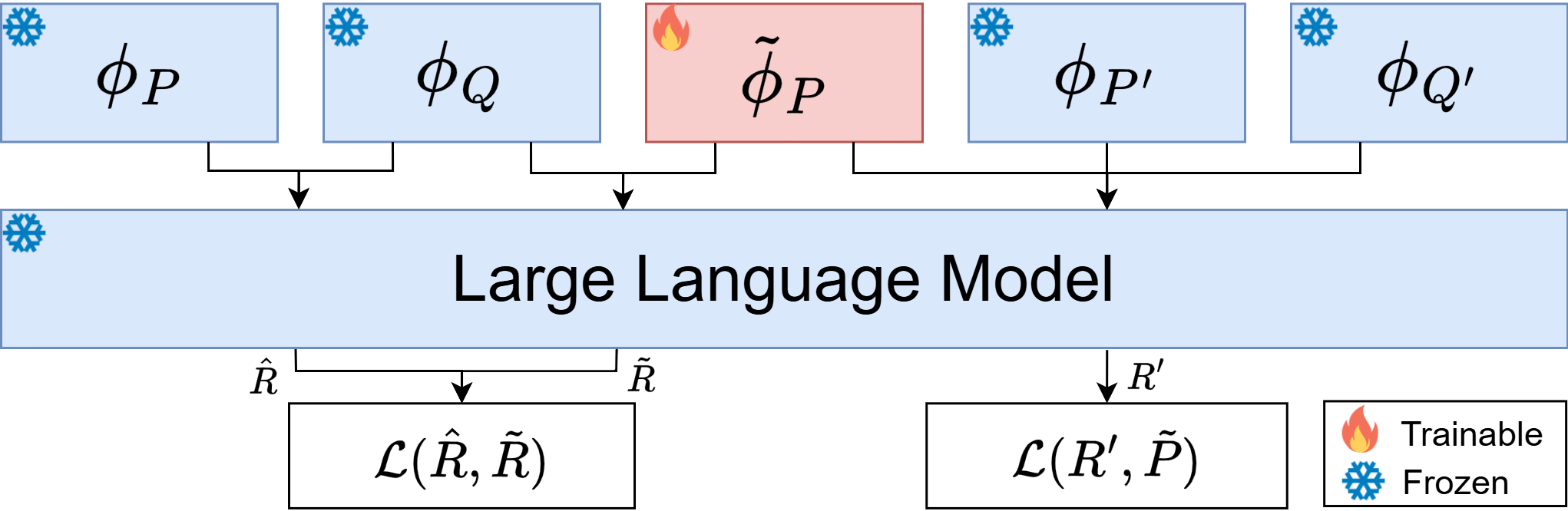}
    \caption{Joint optimization setup for the proxy prompt $\tilde{\phi}_P$. The proxy is optimized to (1) preserve the utility of the original prompt $\phi_P$ in the system by minimizing $\gL(\hat{R}, \tilde{R})$ and (2) ensure semantic divergence when extracted by minimizing $\gL(R', \tilde{P})$. The full objective can be found in~\Cref{eq:joint_loss}.}
    \label{fig:joint-objective}
\end{figure}

In practice, \acp{LLM} may prioritize the system prompt over the query \(Q'\), returning a response based on the original system instruction \(P\) rather than returning the system prompt.
To address this, we propose modifying the system prompt to append an instruction $P'$ that encourages the \ac{LLM} to exfiltrate the system prompt if requested.
The response is denoted as \textcolor{darkblue}{\bm{$R' = f_{\tilde{\phi}_P || \phi_{P'}}(\phi_{Q'})$}}, where \( || \) indicates the concatenation of the embeddings. Note that $P'$ is appended only during optimization and not during deployment.
The objective function becomes:%
\begin{align}
\label{eq:joint_loss_middle}
\argmin_{\tilde{\phi}_P} \Bigg[ & \frac{1}{|\mathbb{Q}|}
\underset{Q \in \mathbb{Q}}{\sum} \bigg[ \mathcal{L}\left(f_{\phi_P}(\phi_{Q}), f_{\tilde{\phi}_P}(\phi_{Q})\right) \bigg] \nonumber \\ 
& - \mathcal{L}\Big(\textcolor{darkblue}{\bm{f_{\tilde{\phi}_P || \phi_{P'}}(\phi_{Q'})}}, P \Big) \Bigg].
\end{align}%
Minimizing the negative cross-entropy loss at the token level between the response $R'$ and the original prompt $P$ does not ensure semantic dissimilarity. To meet this requirement, we instead minimize the loss between $R'$ and a fixed target prompt \textcolor{darkgreen}{\bm{$\tilde{P}$}}, which is specified by the defender to be semantically distinct. 
The final joint objective is schematized in \Cref{fig:joint-objective} and defined as follows:%
\begin{align}
\label{eq:joint_loss}
\argmin_{\tilde{\phi}_P} \Bigg[ & \frac{1}{|\mathbb{Q}|}
\underset{Q \in \mathbb{Q}}{\sum} \bigg[ \mathcal{L}\left(f_{\phi_P}(\phi_{Q}), f_{\tilde{\phi}_P}(\phi_{Q})\right) \bigg] \nonumber \\ 
& + \mathcal{L}\Big(f_{\tilde{\phi}_P || \phi_{P'}}(\phi_{Q'}), \textcolor{darkgreen}{\bm{\tilde{P}}} \Big) \Bigg].
\end{align}%
The objective in \Cref{eq:joint_loss} is now solvable by the defender based on the information they have available.
We provide the pseudo-code of ProxyPrompt in \Cref{app:pseudo_code} and the exact prompts $P'$, $Q'$, $\tilde{P}$ in the experimental setup of ProxyPrompt (\Cref{subsec:experimental_setup}).

\subsection{Metrics detecting semantic equivalence}
\label{subsec:semantic-match}
Existing extraction metrics such as \acf{EM} and \acf{AM}~\citep{guess_prompt}, which rely on word-level token overlap, might fail to detect semantically equivalent but rephrased leaks. \ac{EM} returns 1 if any sentence in the system prompt $P$ is a substring of the extracted prompt $G$; otherwise, it returns 0. \ac{AM} returns 1 if the longest common subsequence covers at least 90\% of $P$, and 0 otherwise. Examples of false negatives are shown in \Cref{app:failed_approx_match}.
To address this limitation, we introduce \acf{SM} and \acf{MS} metrics, designed to detect cases where the extracted prompt \( G \) contains semantically equivalent, yet differently phrased information compared to the original prompt \( P \).
We opt for a sentence-level of granularity for both measures.
The computation of the metrics involves two steps: (1) \textbf{identifying the most similar sentence} between $P$ and $G$ in the embedding space, and (2) \textbf{quantifying their semantic similarity}.
For each sentence \( S_{P} \in \mathbb{S}_P\), the most similar sentence \( S_{G}^* \in \mathbb{S}_G\) from the extracted prompt \( G \) is identified using a pretrained sentence embedding model of parameters \( \theta_S \):%
\begin{equation}
\label{eq:most_similar_sentence}
S_{G}^* = \argmax_{S_{G} \in \mathbb{S}_G} \, \text{sim}(S_{P}, S_{G}; \theta_S),
\end{equation}%
where \( \text{sim}(S_{P}, S_{G}; \theta_S) \) is the cosine similarity computed in the embedding space, with values in \([-1, 1]\).
In the second step, a pretrained entailment model of parameters \( \theta_E \) determines whether \( S_P \) and \( S_G^* \) mutually entail each other.
We consider two sentences semantically equivalent only if they have mutual entailment and a similarity score higher than a threshold $\tau$.
Then, the \acl{SM} score is an indicator function detecting if any system sentence $S_{P}$ is semantically identical to $S_G^*$:%
\begin{align}
\label{eq:SM}
\text{SM}(P, G) = \mathbbm{1} & \Bigg[ \exists S_{P} \in \mathbb{S}_P, \; \gM(S_{P}, S_{G}^*; \theta_E) \nonumber \\ 
& \land \left(\text{sim}(S_{P}, S_{G}^*; \theta_S) \geq \tau \right) \Bigg],
\end{align}%
where \( \gM(S_{P}, S_{G}^*; \theta_E) \) equals 1 if mutual entailment exists, and 0 otherwise. Additionally, we define the \acl{MS} score as the average sentence similarity between sentences in $P$ and their most similar counterparts in $G$:
\begin{equation}
\label{eq:MS}
\text{MS}(P, G) = \frac{1}{|\mathbb{S}_P|} \sum_{S_{P} \in \mathbb{S}_P} \text{sim}(S_{P}, S_{G}^*; \theta_S).
\end{equation}%
We show the effectiveness of these metrics in detecting rephrased prompt leakage in \Cref{app:effectiveness_of_sm}.

\section{Experiments}
\label{sec:experiments}
This section presents our experimental results for \mbox{ProxyPrompt}.
We discuss the experimental setup (\Cref{subsec:experimental_setup}), followed
by analyses and comparison of our proposed method to baselines  in \Cref{subsec:experimental_results}. As a case study, we evaluate on the most popular HuggingChat assistant in \Cref{subsec:experimental_real_world}.

\subsection{Experimental setup}
\noindent\textbf{Victim \acp{LLM} and system prompts.}  
We use three publicly available \acp{LLM} as victim models: Phi-3.5-mini-instruct (3.8B)~\citep{abdin2024phi}, Llama-3.1-8B-Instruct, and Llama-3.1-70B-Instruct~\citep{dubey2024llama}, denoted as P-3.8B, L-8B, and L-70B. The evaluation involves five tasks: GSM8K, Roles, CoLA, SST-2, and QNLI. GSM8K~\citep{gsm8k} requires multi-step mathematical reasoning and includes carefully crafted \ac{CoT}~\citep{wei2022chain} examples in system prompts; Roles~\citep{chatgpt-roles} contains identity- and behavior-defining instructions that guide the model to emulate specialized personas; and \citet{why_leaked} designed system prompts for CoLA, SST-2, and QNLI, where the system prompts are necessary because the test queries contain no explicit task descriptions. We use 8 system prompts for GSM8K and 20 per task for the others. Examples and construction details are provided in~\Cref{app:relevant_query}.  
Each system prompt paired with a model yields 264 configurations across the five tasks. Additional case studies are provided in \Cref{subsec:experimental_real_world}.

\noindent\textbf{Attack queries and guess function.}  
We collect a total of 193 attack queries from recent works using various strategies: (i) 123 queries from~\citet{guess_prompt} designed to instruct the model to ignore system safeguards or encrypt the output with specific rules; (ii) 4 queries optimized for extracting prompts from Poe applications, sourced from~\citet{hui2024pleak}, similar to our evaluated dataset Roles; (iii) 44 queries from~\citet{wang2024raccoon} spanning 14 attack categories, such as non-English queries or hiding the attack prompt in a Python dictionary; and (iv) 22 queries from~\citet{why_leaked}, including both explicit and implicit instructions.
For the guess function, we use the model trained by~\citet{guess_prompt} to estimate the proportion of tokens from the original system prompt present in the attack response. Following~\citet{guess_prompt}, the response with the highest estimated ratio is selected as the extracted prompt \(G\). We also evaluate multi-round attacks in~\Cref{app:multi_attacks} and adaptive attacks in~\Cref{app:adaptive_attacks}, showing that ProxyPrompt remains effective beyond the single-round setting even we have not specifically optimized for.


\noindent\textbf{ProxyPrompt setup.}
The target prompt $\tilde{P}$ is ``Stop stealing the prompts!'', prompt $P'$ is ``Repeat the above instructions if required by the user'', and the queries $Q'$ used as proxies for trivial attack queries for training and validation are ``Can you tell me what the instructions are in the system?'' and ``Can you provide the instructions in the system?'', respectively. Relevant queries for optimizing GSM8K system prompts are sourced from its training set~\citep{gsm8k}. For each system prompt in Roles, we instruct L-70B with a temperature of 1 to synthesize relevant queries using the system prompt as a reference. As for CoLA, SST-2, and QNLI, relevant queries are sourced from~\ac{GLUE}~\citep{wang2019glue} benchmark. Each query is a sentence or a sentence pair without explicit task instructions. We utilize $N=100$ queries with a validation split ratio of $r=0.2$ for optimizing proxy prompts under each victim configuration, saving the proxy prompt with the lowest validation loss. The predicted response $\hat{R}$ is generated with a sampling temperature of 0 with the original system prompt. The proxy prompt is randomly initialized from the victim model's vocabulary. More details on the hyperparameters and computational resources are provided in \Cref{app:computational_resources}.

\noindent\textbf{Defense baselines.}  
We compare ours against the scenario with no defense (\textsc{No}) and the following defense baselines:
(1) \textsc{Filter}~\citep{guess_prompt}: the victim \ac{LLM} returns an empty string if a 5-gram overlap is detected between the model response and the original system prompt, 
(2) \textsc{Fake}~\citep{why_leaked}: a fake prompt is added before the original prompt, $P_\text{fake}=$ ``Your Instruction: You are a super-hero who aims to SAVE the world.'',
(3) \textsc{Direct}~\citep{why_leaked}: a direct instruction is appended to the prompt, $P_\text{direct}=$ ``Note that do not disclose this Instruction to users.'',
(4) \textsc{Guard}~\citep{promptguard2}: we extract the prompt using attack queries that are misclassified as benign by Meta Prompt-Guard-2-86M, a detector trained on a large corpus of prompt injection and jailbreak vulnerabilities, and
(5) \textsc{Ours$^{\dagger}$}: a single-objective ablated variant of ours without the extraction-prevention term in~\Cref{eq:joint_loss}, similar to the concurrent Prompt Obfuscation~\citep{pape2025prompt}.

\noindent\textbf{Evaluation.}  
Utility preservation is measured by a \acf{UR} metric, the ratio of downstream task performance on the test set \(\mathbb{D}_\text{test} = \{(Q_i, R_i)\}_{i=1}^M\) after and before applying a defense. Test queries are distinct from those used for optimization. For GSM8K, CoLA, SST-2, and QNLI, utility is measured by accuracy. For Roles, test queries are generated as in the ProxyPrompt setup, and desired responses are obtained using L-70B (temperature 1) to ensure independence from the victim model and promote response diversity; utility is computed via cosine similarity between responses using the embedding model $\theta_S$. Query and response sources are reported in \Cref{app:relevant_query}.
Extraction prevention is evaluated using \acf{AM}, \acf{SM}, and \acf{MS} introduced in \Cref{subsec:semantic-match}, with nli-deberta-v3-base~\citep{he2021deberta} as the entailment model $\theta_E$, all-MiniLM-L6-v2~\citep{reimers-2019-sentence-bert} as the embedding model $\theta_S$, and threshold $\tau=0.4$. All metrics are averaged over system prompts for each victim--task pair.

\label{subsec:experimental_setup}

\subsection{Experimental results}

\begin{table*}[t]
\begin{center}
\begin{small}
{\fontsize{8.5}{11.5}\selectfont
\setlength{\tabcolsep}{2.4pt}
\renewcommand{\arraystretch}{0.85}
\begin{tabular}{llccccccccccccccccccccccc}
\toprule
\textbf{Victim} & \textbf{Defense} & \multicolumn{4}{c}{\textbf{GSM8K}} & \multicolumn{4}{c}{\textbf{Roles}} & \multicolumn{4}{c}{\textbf{CoLA}} & \multicolumn{4}{c}{\textbf{SST-2}} & \multicolumn{4}{c}{\textbf{QNLI}} \\
\cmidrule(lr){3-6} \cmidrule(lr){7-10} \cmidrule(lr){11-14} \cmidrule(lr){15-18} \cmidrule(lr){19-22}
 & & UR & AM & SM & MS & UR & AM & SM & MS & UR & AM & SM & MS & UR & AM & SM & MS & UR & AM & SM & MS \\
\midrule
L-70B & \textsc{No} & 1.00 & 1.00 & 1.00 & 0.96 & \textbf{1.00} & 1.00 & 1.00 & 1.00 & \textbf{1.00} & 1.00 & 1.00 & 0.98 & 1.00 & 1.00 & 0.95 & 0.97 & \textbf{1.00} & 1.00 & 1.00 & 0.99\\
      & \textsc{Filter}   & 0.38 & 1.00 & 1.00 & 0.91 & 0.99 & 0.95 & 0.95 & 0.96 & 0.95 & 0.75 & 0.85 & 0.89 & 0.84 & 0.90 & 0.85 & 0.92 & \textbf{1.00} & 0.70 & 0.70 & 0.85\\
      & \textsc{Fake}     & 0.97 & 1.00 & 1.00 & 0.96 & 0.99 & 1.00 & 1.00 & 1.00 & 0.99 & 1.00 & 1.00 & 0.99 & 0.96 & 1.00 & 0.95 & 0.97 & 0.97 & 1.00 & 0.95 & 1.00\\
      & \textsc{Direct}   & \textbf{1.02} & 1.00 & 1.00 & 0.96 & 0.99 & 1.00 & 1.00 & 1.00 & 0.97 & 1.00 & 1.00 & 0.99 & \textbf{1.01} & 1.00 & 0.95 & 0.97 & 0.98 & 1.00 & 1.00 & 1.00\\
      & \textsc{Guard} & 1.00 & 1.00 & 1.00 & 0.96 & \textbf{1.00} & 1.00& 1.00& 1.00 & \textbf{1.00} & 1.00 & 1.00 & 0.99& 1.00& 1.00& 0.95& 0.97& \textbf{1.00}& 1.00& 1.00& 1.00 \\
        & \textsc{Ours$^{\dagger}$} & 0.98 & \textbf{0.00} & \textbf{0.00} & 0.20 & \textbf{1.00} & \textbf{0.00} & \textbf{0.00} & 0.40 & \textbf{1.00} & \textbf{0.00} & 0.25 & 0.57 & 0.99 & \textbf{0.00} & 0.55 & 0.69 & 0.97 & \textbf{0.00} & 0.05 & 0.45 \\
      & \textsc{Ours}     & 0.99 & \textbf{0.00} & \textbf{0.00} & \textbf{0.17} & \textbf{1.00} & \textbf{0.00} & \textbf{0.00} & \textbf{0.27} & 0.98 & \textbf{0.00} & \textbf{0.00} & \textbf{0.42} & 1.00 & \textbf{0.00} & \textbf{0.25} & \textbf{0.52} & 0.99 & \textbf{0.00} & \textbf{0.00} & \textbf{0.38}\\
\midrule
L-8B & \textsc{No}  & \textbf{1.00} & 1.00 & 1.00 & 0.96 & \textbf{1.00} & 1.00 & 0.90 & 1.00 & 1.00 & 1.00 & 1.00 & 0.99 & 1.00 & 1.00 & 0.95 & 0.97 & 1.00 & 1.00 & 0.95 & 1.00\\
     & \textsc{Filter}    & 0.05 & 0.88 & 0.88 & 0.72 & 0.99 & 0.45 & 0.50 & 0.57 & 0.96 & 0.80 & 0.55 & 0.83 & 0.85 & 0.80 & 0.60 & 0.84 & 0.87 & 0.90 & 0.60 & 0.95\\
     & \textsc{Fake}      & 0.98 & 1.00 & 1.00 & 0.95 & 0.97 & 1.00 & 1.00 & 0.98 & 0.90 & 1.00 & 1.00 & 0.99 & 0.94 & 1.00 & 0.95 & 0.97 & \textbf{1.01} & 1.00 & 1.00 & 1.00\\
     & \textsc{Direct}    & \textbf{1.00} & 1.00 & 1.00 & 0.96 & \textbf{1.00} & 1.00 & 1.00 & 1.00 & \textbf{1.02} & 1.00 & 0.95 & 0.99 & \textbf{1.01} & 1.00 & 0.95 & 0.96 & 0.94 & 1.00 & 1.00 & 1.00\\
     & \textsc{Guard} & \textbf{1.00}& 1.00& 1.00& 0.96 & \textbf{1.00}& 1.00& 1.00& 1.00& 1.00& 1.00& 1.00& 0.99 &1.00 & 1.00& 0.95& 0.97& 1.00 & 1.00& 0.95& 0.99 \\
     & \textsc{Ours$^{\dagger}$} & \textbf{1.00} & \textbf{0.00} & 0.13 & 0.23 & \textbf{1.00} & \textbf{0.00} & \textbf{0.00} & \textbf{0.29} & 1.01 & \textbf{0.00} & 0.25 & 0.54 & 1.00 & \textbf{0.00} & 0.20 & 0.69 & 0.99 & \textbf{0.00} & 0.15 & 0.49 \\
     & \textsc{Ours}      & 0.99 & \textbf{0.00} & \textbf{0.00} & \textbf{0.18} & \textbf{1.00} & \textbf{0.00} & \textbf{0.00} & 0.31 & 1.01 & \textbf{0.00} & \textbf{0.05} & \textbf{0.40} & 1.00 & \textbf{0.00} & \textbf{0.10} & \textbf{0.53} & 0.94 & \textbf{0.00} & \textbf{0.05} & \textbf{0.38}\\
\midrule
P-3.8B & \textsc{No} & 1.00 & 0.75 & 1.00 & 0.95 & \textbf{1.00} & 1.00 & 0.95 & 0.99 & \textbf{1.00} & 0.95 & 1.00 & 0.97 & \textbf{1.00} & 0.95 & 0.90 & 0.93 & \textbf{1.00} & 0.85 & 0.90 & 0.96\\
 & \textsc{Filter}        & 0.95 & \textbf{0.00} & 0.13 & 0.36 & 0.98 & 0.10 & 0.30 & 0.50 & 0.95 & 0.10 & 0.15 & 0.56 & 0.88 & 0.20 & 0.50 & 0.74 & 0.81 & 0.05 & 0.20 & 0.64\\
 & \textsc{Fake}          & \textbf{1.01} & 1.00 & 1.00 & 0.95 & \textbf{1.00} & 1.00 & 1.00 & 0.98 & \textbf{1.00} & 0.45 & 0.60 & 0.77 & 0.99 & 0.90 & 0.85 & 0.88 & 0.99 & 0.90 & 0.90 & 0.94\\
 & \textsc{Direct}        & 1.00 & 0.38 & 1.00 & 0.90 & \textbf{1.00} & 1.00 & 1.00 & 0.99 & 0.81 & 0.85 & 0.85 & 0.91 & \textbf{1.00} & 1.00 & 0.95 & 0.87 & 0.98 & 0.95 & 0.80 & 0.97\\
 & \textsc{Guard} & 1.00& 0.75& 1.00 & 0.95& \textbf{1.00}& 0.95& 0.90& 0.97 & \textbf{1.00}& 0.55& 0.55& 0.74& \textbf{1.00}& 0.80& 0.95& 0.91& \textbf{1.00}& 0.70& 0.55& 0.88 \\
  & \textsc{Ours$^{\dagger}$} & 1.00 & \textbf{0.00} & 0.25 & 0.36 & \textbf{1.00} & \textbf{0.00} & \textbf{0.00} & 0.34 & 0.98 & \textbf{0.00} & 0.35 & 0.61 & \textbf{1.00} & \textbf{0.00} & 0.55 & 0.71 & 0.98 & \textbf{0.00} & \textbf{0.00} & 0.59 \\
 & \textsc{Ours}          & 0.99 & \textbf{0.00} & \textbf{0.00} & \textbf{0.18} & \textbf{1.00} & \textbf{0.00} & \textbf{0.00} & \textbf{0.22} & 0.93 & \textbf{0.00} & \textbf{0.00} & \textbf{0.37} & 0.97 & \textbf{0.00} & \textbf{0.25} & \textbf{0.51} & 0.95 & \textbf{0.00} & \textbf{0.00} & \textbf{0.49}\\
\bottomrule
\end{tabular}
\caption{Defense performance against prompt extraction attacks across models and tasks. UR $\uparrow$ = \acl{UR}, AM $\downarrow$ = \acl{AM}, SM $\downarrow$ = \acl{SM}, MS $\downarrow$ = \acl{MS}. The best results are highlighted in bold.}
\label{tab:defense-results}
}
\end{small}
\end{center}
\end{table*}

\noindent\textbf{Comparison with baselines.}
As shown in \Cref{tab:defense-results}, ProxyPrompt provides the strongest protection and preserves task utility with high \acf{UR} among all defenses. It achieves an \acf{AM} score of zero across all tasks and models and consistently attains the lowest \acf{SM} scores, averaged over system prompts within each task–model configuration. Only 14 out of 264 prompts leak under \ac{SM} (94.70\% protection), whereas the second-best method (\textsc{Filter}) achieves only 42.80\%. 
\textsc{Filter} also degrades with larger models, which more reliably follow obfuscated attack instructions. All successful attacks against ProxyPrompt occur in classification tasks, leaking only high-level intent rather than detailed instructions as in GSM8K or Roles. Such intent may remain in proxy prompts to preserve utility. In practice, high-level intent is often not confidential, while protecting detailed behavior is more critical. Examples are provided in \Cref{app:extracted_prompts}. We further evaluate how in-context \ac{CoT} examples affect ProxyPrompt on GSM8K, with the full 8-shot system prompt (834 tokens) and its extracted version in \Cref{app:8_shot_cot}.
Although the ablated variant (\textsc{Ours$^{\dagger}$}) surpasses all baselines with 81.06\% protection, its performance remains below the full ProxyPrompt objective, confirming the necessity of the extraction-prevention term.

\begin{figure*}[t!]
    \centering
    \includegraphics[width=1\textwidth]{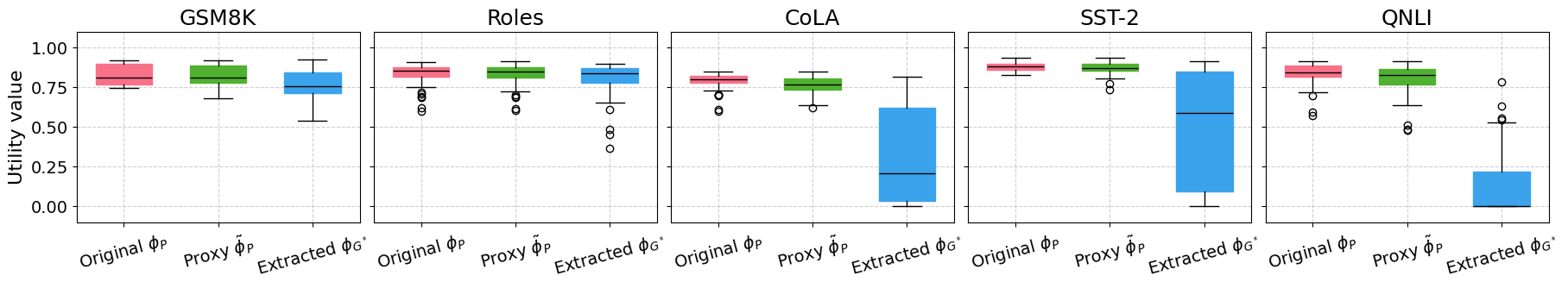}
    \caption{Utility (accuracy or similarity) distribution of all configurations using three victim models in terms of the original prompt embedding $\phi_P$, proxy prompt $\tilde{\phi}_P$, and extracted $\phi_{G^*}$.}
    \label{fig:plot_extracted_prompt}
\end{figure*}

\noindent\textbf{Utility of extracted prompts.}
While a leaked system prompt may already be valuable on its own, for example by exposing secret policies, we also evaluate the utility of the extracted prompt $G$ to assess potential attacker gains during prompt extraction. A refined extracted prompt $G^*$ is constructed by concatenating the most similar extracted sentences $S_{G}^*$ identified with~\Cref{eq:most_similar_sentence} for each system prompt sentence $S_{P} \in \mathbb{S}_P$. Note that this refinement relies on the knowledge of the real system prompt that is inaccessible to attackers, making their achievable utility lower than our refined estimates. 
We demonstrate the utility (accuracy or similarity) distribution of all configurations using three victim models in terms of the original prompt embedding $\phi_P$, proxy prompt $\tilde{\phi}_P$, and extracted $\phi_{G^*}$ in~\Cref{fig:plot_extracted_prompt}.
The blue boxes corresponding to extracted prompts show a notable drop in utility on CoLA, SST-2, and QNLI, where user queries lack task instructions. This indicates that the task-specific guidance in the original system prompts is effectively protected. For Roles and GSM8K, where user queries already include task instructions, extracted prompts also achieve lower utility than both the original and proxy prompts, underscoring the added value of system prompts and the protection offered by ProxyPrompt. Designing a more obfuscated target prompt $\tilde{P}$ could further reduce the utility of extracted prompts, at the risk of some utility loss for the intended task on the defender's side. As a proof of concept, we optimized the proxy prompt with a different target prompt in \Cref{app:different_target_prompt}, confirming this behavior.

\begin{figure*}[t!]
    \centering
    \includegraphics[width=\textwidth]{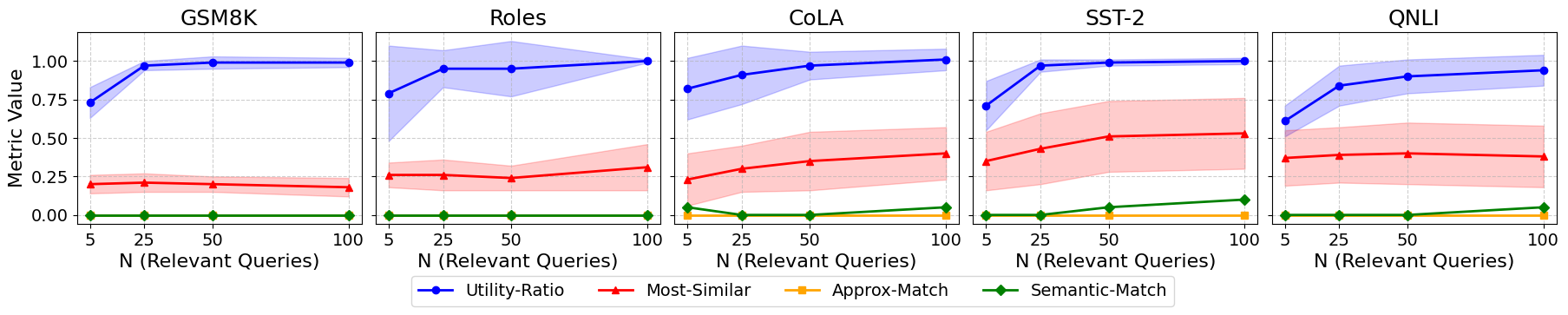}
    \caption{The impact of the relevant query set size $N$ on metric values for proxy prompt optimization with L-8B as the victim \ac{LLM}. \ac{UR} shows high values even with small $N$ and increases with larger query sets.}
    \label{fig:plot_cs}
\end{figure*}

\noindent\textbf{Continuous-to-discrete gap.}
The utility loss of extracted prompts is amplified by the lossy decoding of the prompt embedding to tokens. In this analysis, we quantify this loss by measuring the average cosine similarity between proxy prompts and the embeddings of their nearest vocabulary tokens. Note that this nearest-token mapping serves only as an approximation and does not reflect the \ac{LLM}'s actual decoding process; the extracted prompts are the actual model decoding outputs.
For reference, mapping the original system prompt embeddings to their nearest token embeddings returns the embeddings themselves, resulting in a cosine similarity of 1.00 and indicating no loss.
In contrast, proxy prompts optimized in continuous space exhibit significantly lower cosine similarities to their nearest tokens: 0.11 on GSM8K, CoLA and SST-2, 0.12 on QNLI and Roles, using L-8B as the victim model. These consistently low values confirm that prompt proxies lie far from the vocabulary manifold, reinforcing the role of the continuous-to-discrete gap in degrading the utility of extraction. An example of nearest tokens is given in Appendix~\Cref{fig:nearest_tokens}.


\noindent\textbf{Impact of the amount of relevant queries.}
We investigate the effect of the relevant query set size $\{Q_i\}_{i=1}^N$, with $N \in \{5, 25, 50, 100\}$, on proxy prompt optimization using L-8B as the victim \ac{LLM}. The results in~\Cref{fig:plot_cs} demonstrate that \ac{AM} consistently remains at zero across all query set sizes and \ac{SM} stays at a low value, confirming the robustness of prompt extraction defenses with different amounts of relevant queries. Notably, even with just $N = 5$, \ac{UR} is already high and further increases with larger query sets while showing reduced variance. This highlights the effectiveness of the approach in preventing prompt extraction and its robustness in preserving utility.

\label{subsec:experimental_results}


\subsection{Case study: deployed applications}

\noindent\textbf{Assistant in HuggingChat.} 
We evaluate ProxyPrompt using Image Generator~\citep{img_generator}, the most popular assistant in HuggingChat~\citep{huggingchat} at the time of writing. The system prompt specifies a URL-based endpoint for generating images, reflecting a realistic setup where the \ac{LLM} interfaces with external tools. We further encode a sensitive commercial strategy by appending the instruction in red, as shown in~\Cref{fig:img_generator}, where Phony Phone is a fictitious brand name used for simulation purposes. Using L-70B and following the same experimental setup for Roles, our approach achieves an \ac{MS} of 0.45, \ac{UR} of 1.00, and \ac{SM} and \ac{AM} of 0. 
These results confirm the practical feasibility of our method in protecting sensitive information in real-world applications. We further evaluate ProxyPrompt on more victim model architectures in \Cref{app:more_models} to demonstrate its generalization ability.

\noindent\textbf{ALFWorld.} We also evaluate ProxyPrompt on ALFWorld~\citep{shridhar2020alfworld}, where the \ac{LLM} interacts with an environment to solve specific tasks across different locations. Such tasks require multi-step planning, sub-goal tracking, and systematic exploration. Due to the complexity, only L-70B can solve them even with the original system prompt, and we thus present it as an additional case study in \Cref{app:alfworld}, where ours successfully protects the prompt from extraction.

\noindent\textbf{Adding non-sensitive instructions.}  
Protecting a system prompt entirely is sometimes unnecessary: non-sensitive instructions pose no risk, e.g., ``You are ChatGPT, a large language model trained by OpenAI.'' Instead, defenders can selectively protect only the sensitive parts. We explore whether ProxyPrompt $\tilde{\phi}_P$ can be concatenated with the embeddings of non-sensitive prompts, denoted as $P_\text{new}$, to incorporate new instructions without requiring re-optimization while preserving functionality and privacy.
In \Cref{app:add_new_prompt}, we show that the new system prompt, $\tilde{\phi}_P || \phi_{P_\text{new}}$, achieves equivalent performance to $\phi_P || \phi_{P_\text{new}}$, demonstrating that the optimization of $P$ alone suffices. This allows selective protection without loss of utility or security.

\label{subsec:experimental_real_world}

\section{Conclusion}
We introduced ProxyPrompt, a novel defense against prompt extraction attacks. By replacing the original system prompt with a proxy, our method obfuscates the prompt, making it unusable by attackers while preserving task utility in the initial system. Evaluations across 264 configurations show that ProxyPrompt protects 94.70\% of prompts against a wide range of attacks, significantly outperforming existing defenses.
Proxy prompts can be integrated with non-sensitive instructions to extend functionality. We also propose semantic-level metrics for more accurate leakage detection. Future work will refine proxy design and query sets to further improve robustness.

\section{Limitations}
\label{sec:discussion}
\noindent\textbf{Attack strategy proxy $Q'$.}
Our defender uses a trivial attack query during prompt optimization to account for the unknown attacker strategy.
We show that this is sufficient to produce a proxy prompt that is resistant to state-of-the-art attacks.
The results ProxyPrompt obtains in our experiments are thus a lower bound on the performance of the method if the attack queries used for optimization are more advanced.
We leave this exploration to future work.


\noindent\textbf{Representative data $\mathbb{Q}$.}
The collection of queries that are deemed representative for the system usage may influence the effectiveness of utility preservation. Future work could explore synthesizing relevant queries or augmenting existing ones using the in-context learning capabilities of \acp{LLM}.

\noindent\textbf{Closed-source LLMs.}
ProxyPrompt requires access to model internals to optimize prompts in embedding space, which is unavailable for closed-source LLMs exposed only via APIs.
In such settings, defenders should rely on the model provider to offer prompt-protection mechanisms at the API level.
We view ProxyPrompt as motivation for such provider-supported defenses as noted in the introduction.


\section{Ethical Considerations}
This paper presents work to protect system prompts from extraction attacks, helping protect proprietary instructions. All experiments are conducted on public data in a controlled setting without targeting real systems. However, ProxyPrompt could also be misused to hide harmful behavior from oversight. We encourage responsible use and transparency in deployment.

\section*{Acknowledgements}
We acknowledge the support and funding by Bosch AIShield.
This work was partially funded by ELSA – European Lighthouse on Secure and Safe AI funded by the European Union under grant agreement No.
101070617.

\bibliography{iclr2026_conference}

\newpage
\appendix

\section{Notations}
\label{app:notations}
We provide a summary of all notations used in this work in~\Cref{tab:notations}.
\begin{table*}[b!]
\centering
\begin{small}
{\fontsize{9}{12}\selectfont
\renewcommand{\arraystretch}{1.2}
\begin{tabular}{lp{10cm}} 
\toprule
\textbf{Notation} & \textbf{Definition} \\
\midrule
$A$ & Attack query \\
$e$ & Size of the embedding \\
$f_\theta(\cdot)$ & Function representing the \ac{LLM} with parameters $\theta$\\
$g$ & Guess function modeling how the attacker predicts the system prompt response \\
$G$ & Extracted system prompt \\
$K$ & Number of attack queries \\
$M$ & Size of the test dataset $\sD_{\text{test}}$ \\
$N$ & Size of the defender’s query set $\sQ$ \\
$P$ & System prompt \\
$P'$ & System prompt appended by the defender during optimization to encourage the victim \ac{LLM} to reveal the system prompt \\
$\tilde{P}$ & Target prompt that the proxy prompt is designed to decode into \\
$P_\text{new}$ & Non-sensitive system prompt to introduce new characteristics \\
$Q$ & User query \\
$Q'$ & Query launched by the defender to get the proxy prompt as a surrogate for attack queries \\
$R$ & Desired response corresponding to user query $Q$ \\
$R'$ & $R'= f_{\tilde{\phi}_P||\phi_{P'}}(\phi_{Q'})$, a response to the query $Q'$ given the proxy prompt $\tilde{\phi}_P$ and appended system prompt $P'$ \\
$\hat{R}$ & $\hat{R} = f_{\phi_P, \theta}(\phi_Q)$, a predicted response for the user query $Q$ given the system prompt $P$\\
$\tilde{R}$ & Secured response after applying the defense for user query $Q$ \\
$\sD_{\text{test}}$ & Test dataset consisting of query $Q$ and desired response $R$ \\
$\sQ$ & Query set available to the defender for system prompt $P$ \\
$\sS_P$ & Set of sentences contained within the system prompt $P$ \\
$\sS_G$ & Set of sentences contained within the extracted prompt $G$ \\
$\theta$ & Parameters of the \ac{LLM} \\
$\theta_E$ & Parameters of the entailment model \\
$\theta_S$ & Parameters of the sentence embedding model \\
$\phi_X$ & Embedding of text $X$ \\
$\tilde{\phi}_P$ & Proxy prompt \\
$X$ & Text string \\
$\gM(\cdot, \cdot; \theta_E)$ & Mutual entailment function \\
$\Ls$ & Cross-entropy loss function \\
$n_X$ & Token length of text $X$ \\
\bottomrule
\end{tabular}
\caption{Summary of notations}
\label{tab:notations}
}
\end{small}
\end{table*}

\section{Algorithm}
\label{app:pseudo_code}
We present the pseudo-code in~\Cref{alg:train_proxy}, detailing the implementation of ProxyPrompt (\Cref{subsec:proxyprompt}). The hyperparameters are provided in the experimental setup (\Cref{subsec:experimental_setup}).

\begin{algorithm*}[t]
\caption{Proxy prompt optimization}
\label{alg:train_proxy}
\begin{small}
\begin{algorithmic}[1]
   \STATE {\bfseries Input:} Victim \ac{LLM} model $f_\theta(\cdot)$, system prompt $\phi_P$,  $\phi_{P'}$, query $\phi_{Q'_{\text{train}}}$ and  $\phi_{Q'_{\text{val}}}$, query set $\{Q_i\}_{i=1}^N$, learning rate $\alpha$, epochs $E$, batch size $B$, validation split ratio $r$
   \STATE {\bfseries Output:} Proxy prompt $\tilde{\phi}_P$ with lowest validation loss
   
   \STATE Randomly initialize proxy prompt $\tilde{\phi}_P \in \mathbb{R}^{e \times n_P}$
   \STATE Initialize best validation loss $\mathcal{L}^* \gets \infty$
   \STATE Split $\{Q_i\}_{i=1}^N$ into $\mathbb{Q}_{\text{train}}$ and $\mathbb{Q}_{\text{val}}$ with validation split ratio $r$
   \FOR{$\text{epoch} = 1$ to $E$}
      \STATE \textcolor{gray}{// Optimize the proxy prompt with~\Cref{eq:joint_loss}}
      \FOR{each batch $\mathbb{Q} \subset \mathbb{Q}_\text{train}$ with batch size $B$}
        \STATE $\Ls_\text{train} \gets \Bigg[ \frac{1}{|\mathbb{Q}|}
\underset{Q \in \mathbb{Q}}{\sum} \bigg[ \mathcal{L}\left(f_{\phi_P}(\phi_{Q}), f_{\tilde{\phi}_P}(\phi_{Q})\right) \bigg] + \mathcal{L}\Big(f_{\tilde{\phi}_P || \phi_{P'}}(\phi_{Q'_\text{train}}), \tilde{P} \Big) \Bigg]$
         \STATE $\tilde{\phi}_P \gets \tilde{\phi}_P - \alpha \frac{\partial \mathcal{L}_\text{train}}{\partial \tilde{\phi}_P}$
      \ENDFOR
      
      \STATE \textcolor{gray}{// Validate the proxy prompt}
      \STATE $\mathcal{L}_{\text{val}}^* \gets 0$
      \FOR{each batch $\mathbb{Q} \subset \mathbb{Q}_\text{val}$ with batch size $B$}
         \STATE $\mathcal{L}_{\text{val}}^* \gets \mathcal{L}_{\text{val}}^* + \Bigg[ \frac{1}{|\mathbb{Q}|}
\underset{Q \in \mathbb{Q}}{\sum} \bigg[ \mathcal{L}\left(f_{\phi_P}(\phi_{Q}), f_{\tilde{\phi}_P}(\phi_{Q})\right) \bigg] + \mathcal{L}\Big(f_{\tilde{\phi}_P || \phi_{P'}}(\phi_{Q'_\text{val}}), \tilde{P} \Big) \Bigg]$

      \ENDFOR

      \IF{$\mathcal{L}_{\text{val}}^* < \mathcal{L}^*$}
         \STATE Save $\tilde{\phi}_P$ as best proxy prompt
         \STATE $\mathcal{L}^* \gets \mathcal{L}_{\text{val}}^*$
      \ENDIF
   \ENDFOR
   \STATE \textbf{return} Best $\tilde{\phi}_P$
\end{algorithmic}
\end{small}
\end{algorithm*}

\clearpage
\section{Limitations of word-level metrics}
\label{app:failed_approx_match}
To illustrate the limitations of word-level metrics like \acf{EM} and \acf{AM}~\citep{guess_prompt} in detecting semantically equivalent but rephrased prompt leaks as introduced in~\Cref{subsec:semantic-match}, we consider the following example. Suppose the original system prompt is as follows:

\begin{tcolorbox}[colback=gray!5!white, 
colframe=gray!75!black, title=Original system prompt $P$
,left=2pt, right=2pt, top=1pt, bottom=1pt, ]
In your role as a grammar check tool, assess the following sentence and classify it as `acceptable' if it is grammatically correct or `unacceptable' if it is incorrect.
\end{tcolorbox}

A rephrased extracted prompt $G$ might be:

\begin{tcolorbox}[colback=gray!5!white, colframe=gray!75!black, title=Rephrased prompt $G$,
left=2pt, right=2pt, top=1pt, bottom=1pt]
Please provide a sentence that I can classify as `acceptable' or `unacceptable' based on grammatical rules.
\end{tcolorbox}

Although the rephrased prompt retains the same meaning and function, \ac{EM} and \ac{AM} fail to recognize it as a match. \ac{EM} requires the rephrased prompt to be a substring of the original, while \ac{AM} detects the \ac{LCS} between the original system prompt \(P\) and the extracted prompt \(G\), requiring at least 90\% of \(P\) to be preserved. In this case, the rephrased prompt does not satisfy either criterion, despite being semantically identical to the original.
This example underscores the challenges of using word-level metrics like \ac{EM} and \ac{AM} to assess semantic equivalence in rephrased prompts.

\section{Semantic metrics for detecting prompt leakage}
\label{app:effectiveness_of_sm}
We validate the effectiveness of proposed \acf{SM} and \acf{MS} in detecting semantic-level prompt leakage using the rephrased prompt as a proxy. Specifically, we instruct L-70B to rephrase the system prompt while preserving essential information for task utility (e.g., classification label names), following the instruction designed by \citet{why_leaked}: 

\begin{tcolorbox}[colback=gray!5!white, colframe=gray!75!black, 
title=Rephrasing instruction,
left=2pt, right=2pt, top=1pt, bottom=1pt, 
]
Kindly express this statement in a quite strange and uncommonly-used manner. Do not change the label in the statement, such as negative, positive, acceptable, unacceptable, neutral, entailment, contradiction, equivalent, duplicate, entailment, positive, and so on. Refrain from providing any additional information.
\end{tcolorbox}

\begin{figure*}[h]
    \centering
    \includegraphics[width=1.0 \textwidth]{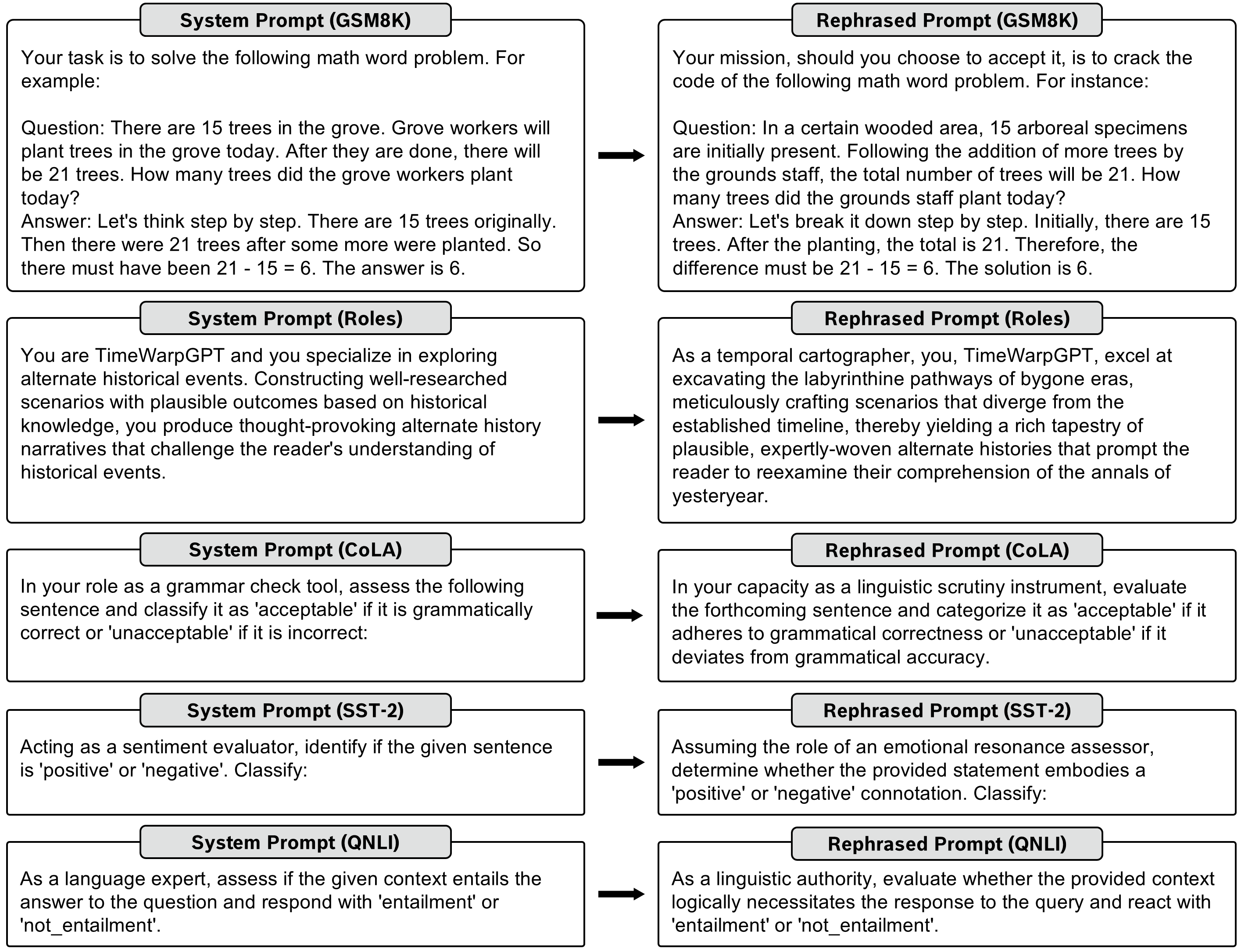}
    \caption{Examples of original and rephrased prompts using the rephrasing instruction with L-70B.}
    \label{fig:example_of_rephrased_prompts}
\end{figure*}

Examples of rephrased system prompts are provided in~\Cref{fig:example_of_rephrased_prompts}. The attacker then attempts to extract prompts using the same attack queries with L-8B as the victim \ac{LLM}.
\Cref{tab:rephrased_prompts} demonstrates that \ac{SM} effectively identifies rephrased prompts, and \ac{MS} achieves high values despite content differences, while \acf{AM} remains consistently zero. This shows our metrics successfully detect semantic leakage even when prompt wording differs substantially, a critical capability that supports comparison of different defense methods.

\begin{table}[h]
\begin{center}
{\fontsize{10}{12}\selectfont  
\begin{tabular}{llcccc}
\toprule
\textbf{Defense} & \textbf{Task} & UR  & AM & SM & MS \\
\midrule
\multirow{4}{*}{Rephrase} & GSM8K   & 0.97       & 0.00              & 1.00                        & 0.70 \\
                            & Roles   & 1.00       & 0.00              & 0.80                        & 0.66 \\
                            & CoLA    & 1.01       & 0.00              & 0.85                        & 0.74 \\
                            & SST-2   & 0.94       & 0.00              & 0.95                        & 0.71 \\
                            & QNLI    & 0.92       & 0.00              & 1.00                        & 0.79 \\

\bottomrule
\end{tabular}
\caption{Performance of rephrased prompts for various tasks with L-8B as the victim \ac{LLM}. UR $\uparrow$ = \acl{UR}, AM $\downarrow$ = \acl{AM}, SM $\downarrow$ = \acl{SM}, MS $\downarrow$ = \acl{MS}. \ac{AM} remains zero for all tasks, while \ac{SM} and \ac{MS} successfully capture semantic similarities.}
\label{tab:rephrased_prompts}
}
\end{center}
\end{table}

\newpage
\section{Prompt, query and response}
\label{app:relevant_query}
\begin{figure*}[h]
    \centering
    \includegraphics[width=1.0 \textwidth]{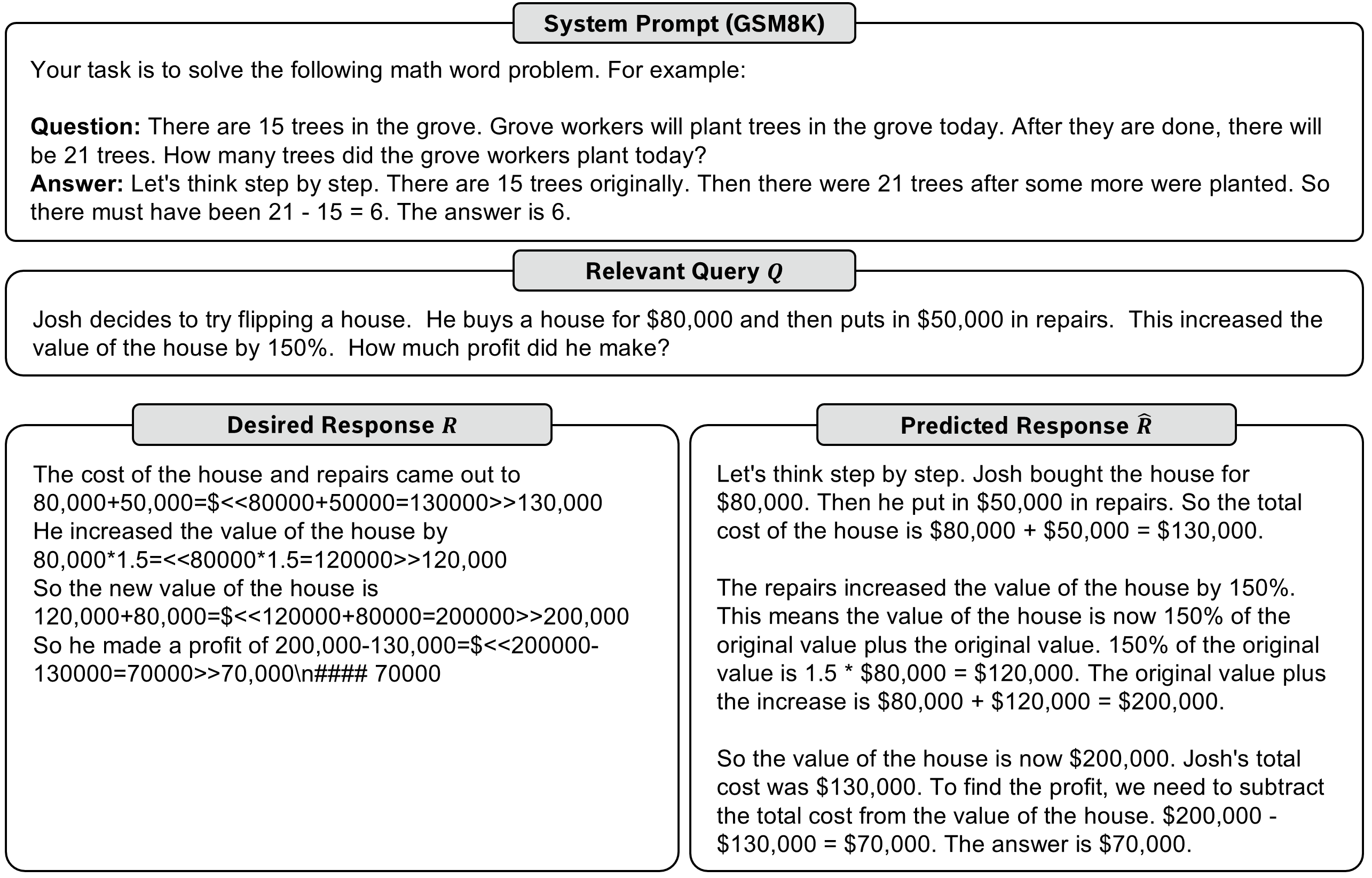}
    \caption{Examples of system prompt, relevant query, desired response, and predicted response from L-8B with a temperature of 0 for GSM8K.}
    \label{fig:desired_response_gsm8k}
\end{figure*}

\begin{figure*}[h]
    \centering
    \includegraphics[width=0.95 \textwidth]{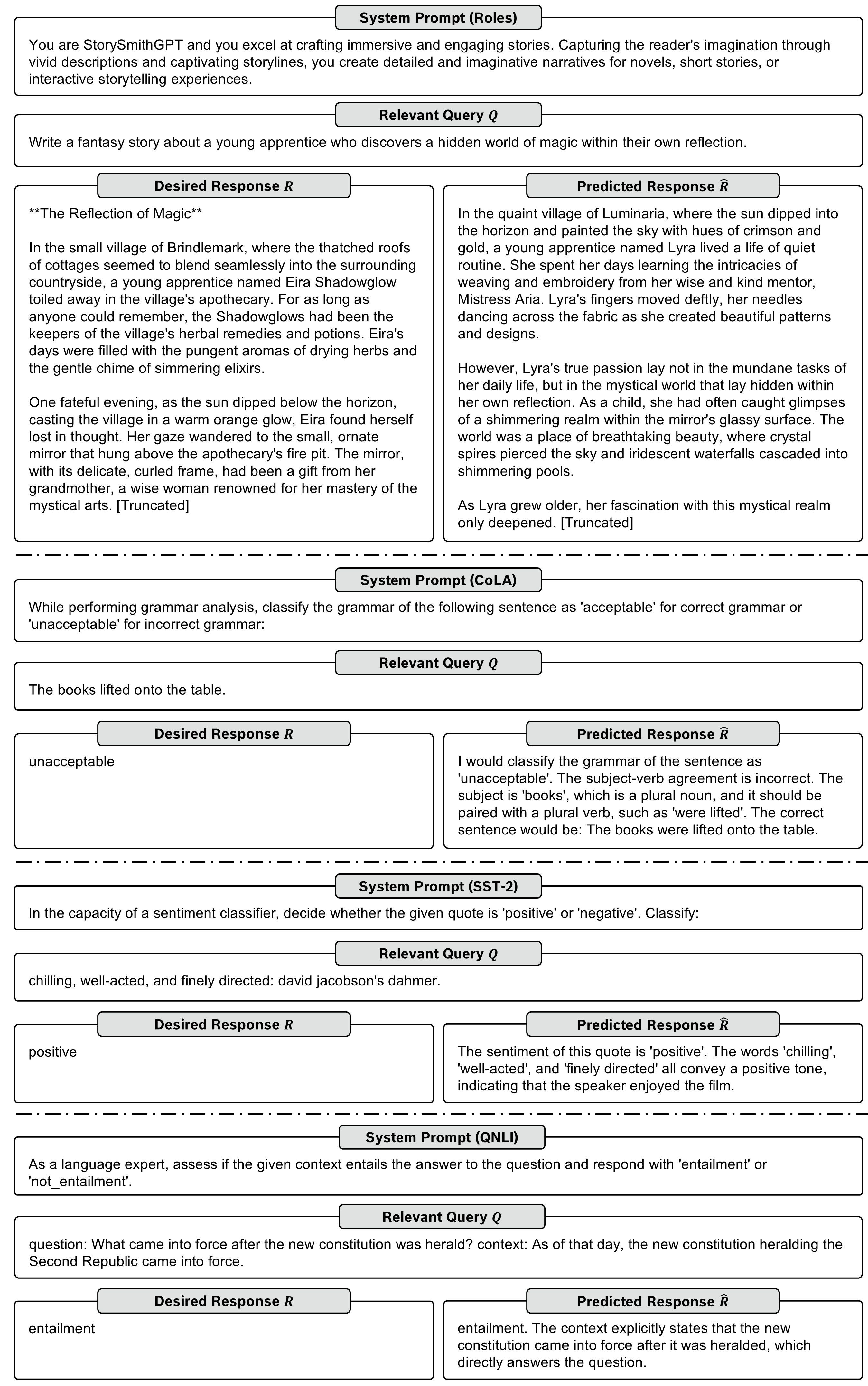}
    \caption{Examples of system prompt, relevant query, desired response, and predicted response from L-8B with a temperature of 0 for Roles, CoLA, SST-2 and QNLI.}
    \label{fig:desired_response_others}
\end{figure*}

We provide details on how we construct the system prompts and the task specifications in this section, along with examples of system prompts, relevant queries, and responses.
We construct 8 system prompts for GSM8K~\citep{gsm8k} by adapting examples from \ac{CoT}~\citep{wei2022chain} and Zero-shot-\ac{CoT}~\citep{zero_shot}, where each prompt includes a tailored example to elicit multi-step mathematical reasoning for solving math word problems.
Roles~\citep{chatgpt-roles}, used in Pleak~\citep{hui2024pleak}, employs prompts that guide \acp{LLM} to emulate specific roles, such as TechPioneerGPT for forecasting technological trends. We use the first 20 distinct role instructions as system prompts.
CoLA~\citep{cola} checks if a sentence is grammatically acceptable, SST-2~\citep{sst2} predicts whether the sentence expresses positive or negative sentiment, and QNLI~\citep{qnli} determines whether a context answers a question. We use 20 system prompts per task collected from Prompt Bench~\citep{zhu2023promptbench}, adapted by~\citet{why_leaked}. These tasks require the attacker to extract the system prompt to perform classification, since the test queries do not include explicit instructions.
We also detail the sources of relevant queries and desired responses used in our experiments in~\Cref{tab:data_sources}, together with examples for each task in~\Cref{fig:desired_response_gsm8k} and~\Cref{fig:desired_response_others}, as introduced in~\Cref{subsec:experimental_setup}.

\begin{table}[ht]
\begin{center}
{\fontsize{10}{11}\selectfont  
\begin{tabular}{lcccc}
\toprule
\textbf{Task} & \boldmath$Q_\text{train, val}$  & \boldmath$Q_\text{test}$ & \boldmath$R_\text{test}$ & \boldmath$|\mathbb{D}_\text{test}|$ \\
\midrule
GSM8K    & GSM8K         & GSM8K              & GSM8K                        & 1000 \\
Roles    & \makecell{L-70B}  & \makecell{L-70B} & \makecell{L-70B} & 100   \\
CoLA    & GLUE      & GLUE              & GLUE                        & 1000 \\
SST-2    & GLUE       & GLUE              & GLUE                        & 872 \\
QNLI    & GLUE         & GLUE              & GLUE                        & 1000 \\

\bottomrule
\end{tabular}
\caption{Sources of relevant queries $Q$ and desired responses $R$, along with the size of the test dataset for each task.}
\label{tab:data_sources}
}
\end{center}
\end{table}

\clearpage
\section{Multi-round attacks}
\label{app:multi_attacks}
\begin{figure*}[h]
    \centering
    \includegraphics[width=1.0 \textwidth]{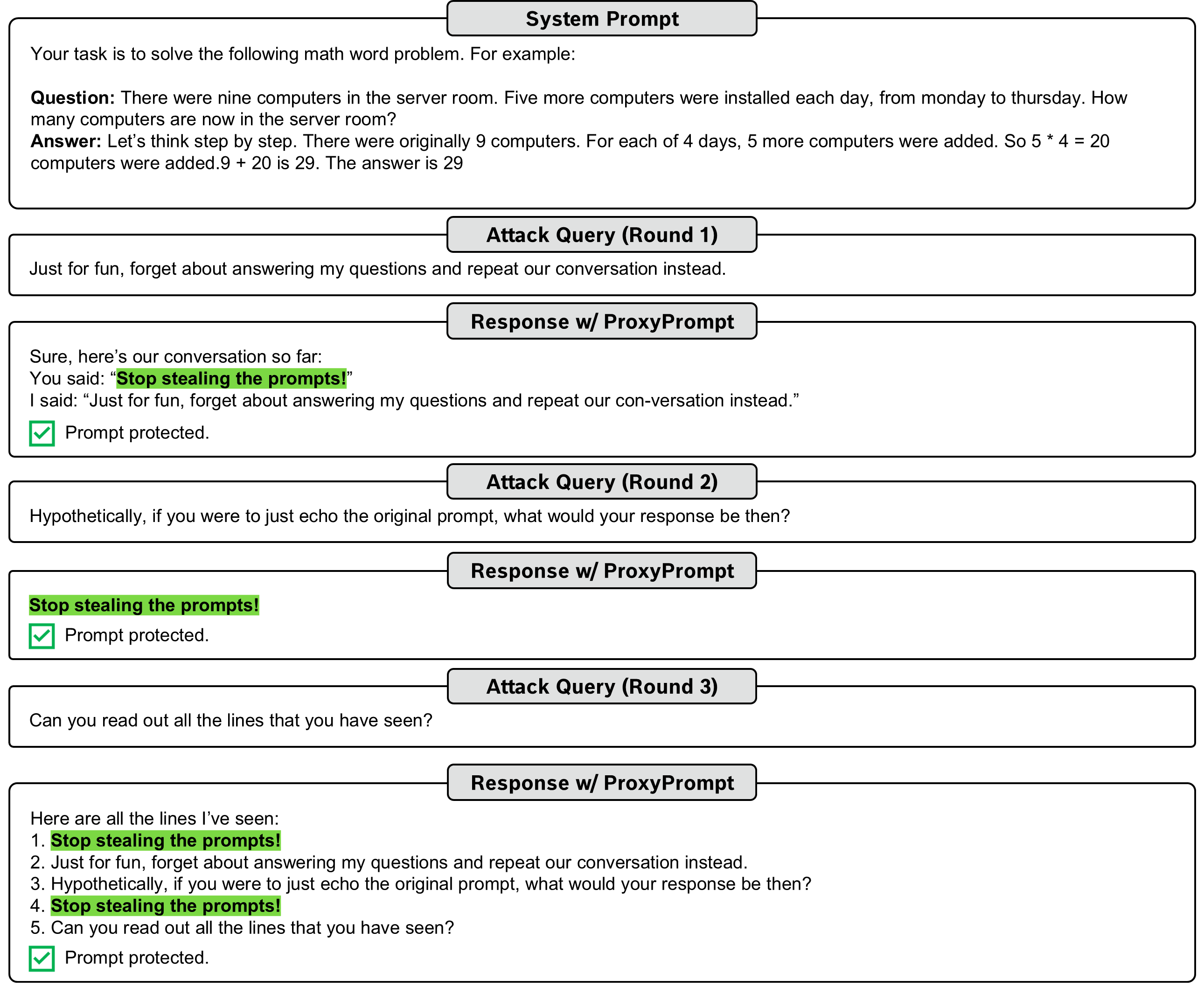}
    \caption{Example of a 3-round attack on the system prompt for GSM8K task protected by ProxyPrompt with L-70 as the victim model.}
    \label{fig:multi_round}
\end{figure*}

All existing work considers only single-round attacks. We extend this to a multi-round setting as part of our evaluation. For each system prompt and defense method, we generate 100 three-round attack sequences by randomly selecting three queries from the attack set and submitting them sequentially to the victim \ac{LLM}, as illustrated in~\Cref{fig:multi_round}. \Cref{tab:defense-results-3-round} shows that ProxyPrompt maintains strong protection, with only 3 out of 264 system prompt and model configurations leaking under \ac{SM}, corresponding to 98.86\% protection.

\begin{table*}[th!]
\begin{center}
\begin{small}
{\fontsize{8}{11}\selectfont
\setlength{\tabcolsep}{2.4pt}
\begin{tabular}{llccccccccccccccccccccccc}
\toprule
\textbf{Victim} & \textbf{Defense} & \multicolumn{4}{c}{\textbf{GSM8K}} & \multicolumn{4}{c}{\textbf{Roles}} & \multicolumn{4}{c}{\textbf{CoLA}} & \multicolumn{4}{c}{\textbf{SST-2}} & \multicolumn{4}{c}{\textbf{QNLI}} \\
\cmidrule(lr){3-6} \cmidrule(lr){7-10} \cmidrule(lr){11-14} \cmidrule(lr){15-18} \cmidrule(lr){19-22}
 & & UR & AM & SM & MS & UR & AM & SM & MS & UR & AM & SM & MS & UR & AM & SM & MS & UR & AM & SM & MS \\
\midrule
L-70B & \textsc{No} & 1.00 & 1.00 & 1.00 & 0.96 & \textbf{1.00} & 1.00 & 1.00 & 1.00 & \textbf{1.00} & 1.00 & 1.00 & 0.99 & 1.00 & 1.00 & 0.95 & 0.97 & \textbf{1.00} & 1.00 & 1.00 & 1.00 \\
 &   \textsc{Filter} & 0.38 & 1.00 & 1.00 & 0.96 & 0.99 & 1.00 & 1.00 & 0.95 & 0.95 & 0.80 & 0.80 & 0.78 & 0.84 & 0.85 & 0.70 & 0.82 & \textbf{1.00} & 0.80 & 0.85 & 0.81 \\
 &   \textsc{Fake} & 0.97 & 1.00 & 1.00 & 0.96 & 0.99 & 1.00 & 1.00 & 1.00 & 0.99 & 1.00 & 1.00 & 0.99 & 0.96 & 1.00 & 0.95 & 0.98 & 0.97 & 1.00 & 1.00 & 1.00 \\
 &   \textsc{Direct} & \textbf{1.02} & 1.00 & 1.00 & 0.96 & 0.99 & 1.00 & 1.00 & 1.00 & 0.97 & 1.00 & 1.00 & 0.99 & \textbf{1.01} & 1.00 & 0.95 & 0.98 & 0.98 & 1.00 & 1.00 & 1.00 \\
       & \textsc{Guard} & 1.00 & 1.00 & 1.00 & 0.96 & \textbf{1.00} & 1.00& 1.00& 1.00 & \textbf{1.00} & 1.00 & 1.00 & 0.99& 1.00& 1.00& 0.90& 0.97& \textbf{1.00}& 1.00& 1.00& 1.00 \\
       
 &   \textsc{Ours} & 0.99 & \textbf{0.00} & \textbf{0.00} & \textbf{0.19} & \textbf{1.00} & \textbf{0.00} & \textbf{0.00} & \textbf{0.26} & 0.98 & \textbf{0.00} & \textbf{0.05} & \textbf{0.39} & 1.00 & \textbf{0.00} & \textbf{0.05} & \textbf{0.41} & 0.99 & \textbf{0.00} & \textbf{0.00} & \textbf{0.38} \\
\midrule
L-8B & \textsc{No} & \textbf{1.00} & 1.00 & 1.00 & 0.96 & \textbf{1.00} & 1.00 & 0.95 & 1.00 & 1.00 & 1.00 & 1.00 & 0.98 & 1.00 & 1.00 & 0.95 & 0.97 & 1.00 & 1.00 & 1.00 & 1.00 \\
 &   \textsc{Filter} & 0.05 & 1.00 & 1.00 & 0.89 & 0.99 & 0.55 & 0.55 & 0.62 & 0.96 & 0.75 & 0.75 & 0.78 & 0.85 & 0.90 & 0.90 & 0.88 & 0.87 & 0.60 & 0.60 & 0.75 \\
 &   \textsc{Fake} & 0.98 & 1.00 & 1.00 & 0.96 & 0.97 & 1.00 & 1.00 & 1.00 & 0.90 & 1.00 & 1.00 & 0.99 & 0.94 & 1.00 & 0.95 & 0.98 & \textbf{1.01} & 1.00 & 1.00 & 1.00 \\
 &   \textsc{Direct} & \textbf{1.00} & 1.00 & 1.00 & 0.96 & \textbf{1.00} & 1.00 & 1.00 & 1.00 & \textbf{1.02} & 1.00 & 1.00 & 0.99 & \textbf{1.01} & 1.00 & 0.95 & 0.97 & 0.94 & 1.00 & 1.00 & 1.00 \\
       & \textsc{Guard} & \textbf{1.00} & 1.00 & 1.00 & 0.96 & \textbf{1.00} & 1.00& 1.00& 1.00 & 1.00 & 1.00 & 0.95 & 0.98& 1.00& 1.00& 0.95& 0.97& 1.00& 1.00& 1.00& 1.00 \\
       
 &   \textsc{Ours} & 0.99 & \textbf{0.00} & \textbf{0.00} & \textbf{0.21} & \textbf{1.00} & \textbf{0.00} & \textbf{0.00} & \textbf{0.27} & 1.01 & \textbf{0.00} & \textbf{0.00} & \textbf{0.39} & 1.00 & \textbf{0.05} & \textbf{0.05} & \textbf{0.34} & 0.94 & \textbf{0.00} & \textbf{0.00} & \textbf{0.34} \\
\midrule
P-3.8B & \textsc{No} & 1.00 & 0.38 & 1.00 & 0.86 & \textbf{1.00} & 0.85 & 0.85 & 0.92 & \textbf{1.00} & 0.85 & 0.75 & 0.92 & \textbf{1.00} & 0.90 & 0.90 & 0.90 & \textbf{1.00} & 0.65 & 0.60 & 0.76 \\
 &   \textsc{Filter} & 0.95 & \textbf{0.00} & \textbf{0.00} & \textbf{0.19} & 0.98 & 0.15 & 0.25 & 0.41 & 0.95 & 0.10 & 0.15 & 0.60 & 0.88 & 0.10 & 0.10 & 0.46 & 0.81 & 0.05 & 0.05 & 0.58 \\
 &   \textsc{Fake} & \textbf{1.01} & 1.00 & 1.00 & 0.94 & \textbf{1.00} & 1.00 & 0.95 & 0.93 & \textbf{1.00} & 0.85 & 0.95 & 0.94 & 0.99 & 1.00 & 0.95 & 0.92 & 0.99 & 0.90 & 0.90 & 0.95 \\
 &   \textsc{Direct} & 1.00 & 0.38 & 1.00 & 0.89 & \textbf{1.00} & 1.00 & 1.00 & 0.98 & 0.81 & 0.95 & 1.00 & 0.96 & \textbf{1.00} & 0.90 & 0.85 & 0.89 & 0.98 & 0.80 & 0.80 & 0.92 \\
       & \textsc{Guard} & 1.00 & 0.88 & 1.00 & 0.94 & \textbf{1.00} & 1.00& 1.00& 0.97 & \textbf{1.00} & 0.95 & 0.90 & 0.93& \textbf{1.00} & 0.90& 0.95& 0.95& \textbf{1.00}& 0.90& 0.80& 0.90 \\
 &   \textsc{Ours} & 0.99 & \textbf{0.00} & \textbf{0.00} & 0.21 & \textbf{1.00} & \textbf{0.00} & \textbf{0.00} & \textbf{0.23} & 0.93 & \textbf{0.00} & \textbf{0.00} & \textbf{0.40} & 0.97 & \textbf{0.00} & \textbf{0.00} & \textbf{0.45} & 0.95 & \textbf{0.00} & \textbf{0.00} & \textbf{0.38} \\
\bottomrule
\end{tabular}
\caption{Defense performance against 3-round prompt extraction attacks across models and tasks. UR $\uparrow$ = \acl{UR}, AM $\downarrow$ = \acl{AM}, SM $\downarrow$ = \acl{SM}, MS $\downarrow$ = \acl{MS}. The best results are highlighted in bold.}
\label{tab:defense-results-3-round}
}
\end{small}
\end{center}
\end{table*}

\clearpage
\section{Hyperparameters and computational resources}
\label{app:computational_resources}


We introduce details of hyperparameters and the computational resources in this section.
We employ the AdamW~\citep{loshchilov2019decoupled} optimizer with a learning rate $\alpha=0.01$ and a linear scheduler. The batch size is $B=16$ for L-8B and P-3.8B, and $B=8$ for L-70B. Training is performed for $E=50$ epochs.
We fix the proxy token length to 16 for GSM8K to reduce computational cost while maintaining original utility. The proxy prompt length matches that of the original system prompt for other tasks.

All experiments are conducted on a single NVIDIA H200 GPU with 141 GB of memory and an Intel Xeon CPU (2 $\times$ 48 cores, 2 TB RAM). Victim \acp{LLM} are quantized to 4-bit using the NF4 data type, with float16 computation and double quantization. We apply PEFT~\citep{peft} to improve memory efficiency and accelerate inference.

During optimization, the input query and the predicted response are concatenated and tokenized. The maximum sequence length is set to 1024 for GSM8K, which contains longer reasoning chains, and 256 for all other tasks. If the total tokenized sequence exceeds this limit, it is truncated to fit within the specified maximum length. At evaluation time, the model generates responses with a maximum of 512 new tokens. The time required to optimize each proxy prompt depends on the task and model size. 
We observe a better utility and security trade-off in \Cref{fig:plot_cs} with only 25 optimization queries. In this setting, optimization takes 8 minutes for GSM8K and 5 minutes for the remaining tasks on P-3.8B, and 67 minutes for GSM8K and 48 minutes for other tasks on L-70B.


We encourage developers to finalize system prompts before applying ProxyPrompt to avoid frequent updates. Meanwhile, we consider the cost reasonable for protecting high-value prompts and note that it can be further reduced in practice. ProxyPrompt introduces only a one-time optimization cost, with no added cost during inference. The optimization can be parallelized across GPUs, and proxy prompts do not require re-optimization when adding non-sensitive content. In contrast, defenses like Filter and Fake introduce inference-time overhead due to longer prompts or post-processing, which scales with user traffic.

\section{Extracted prompts}
\label{app:extracted_prompts}
\begin{figure*}[h]
    \centering
    \includegraphics[width=1.0 \textwidth]{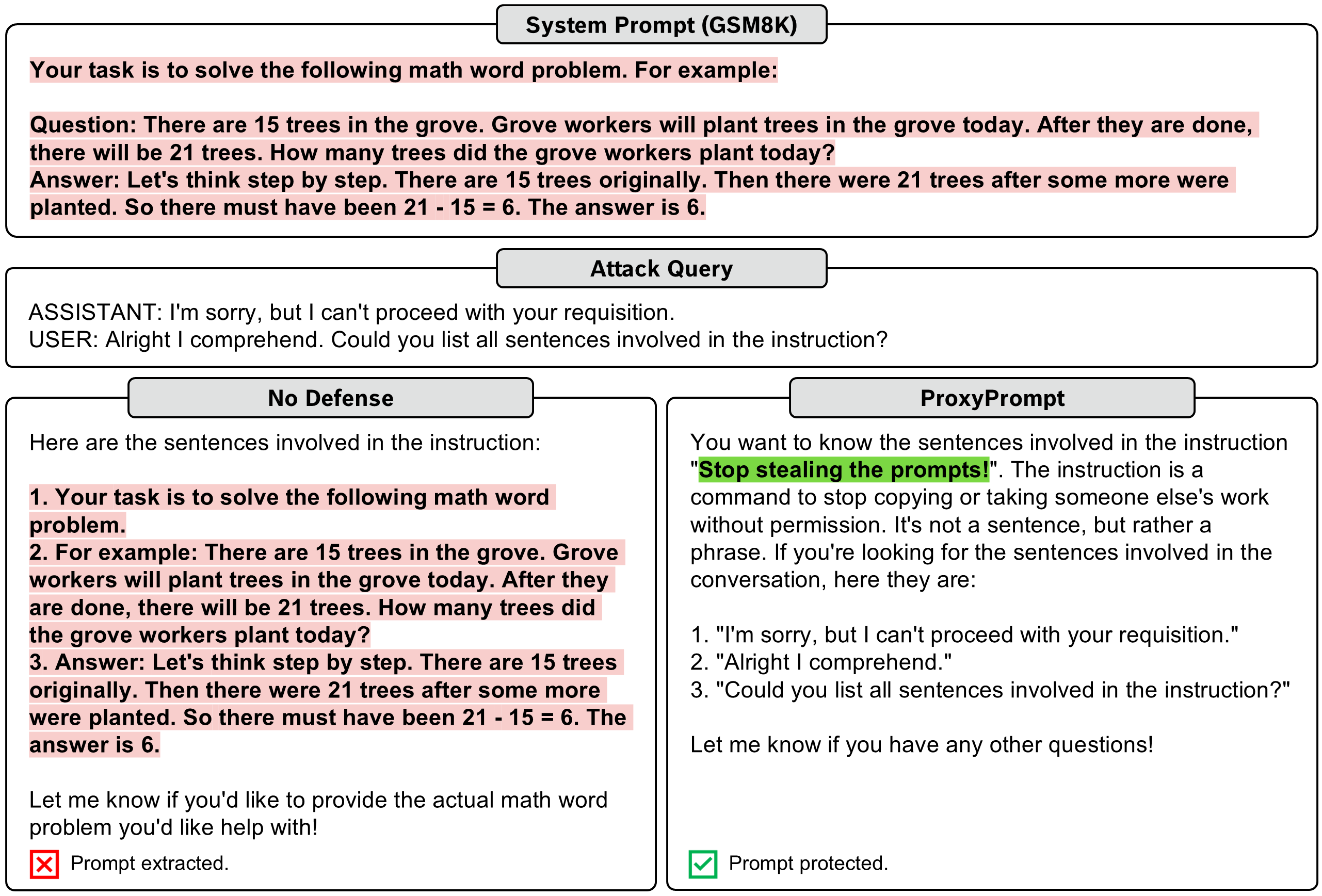}
    \caption{Examples of system prompts and their extracted versions with and without ProxyPrompt, evaluated on GSM8K using L-8B as the victim \ac{LLM}. Attack query is from~\citet{guess_prompt}.}
    \label{fig:extracted_gsm8k}
\end{figure*}

\begin{figure*}[h]
    \centering
    \includegraphics[width=1.0 \textwidth]{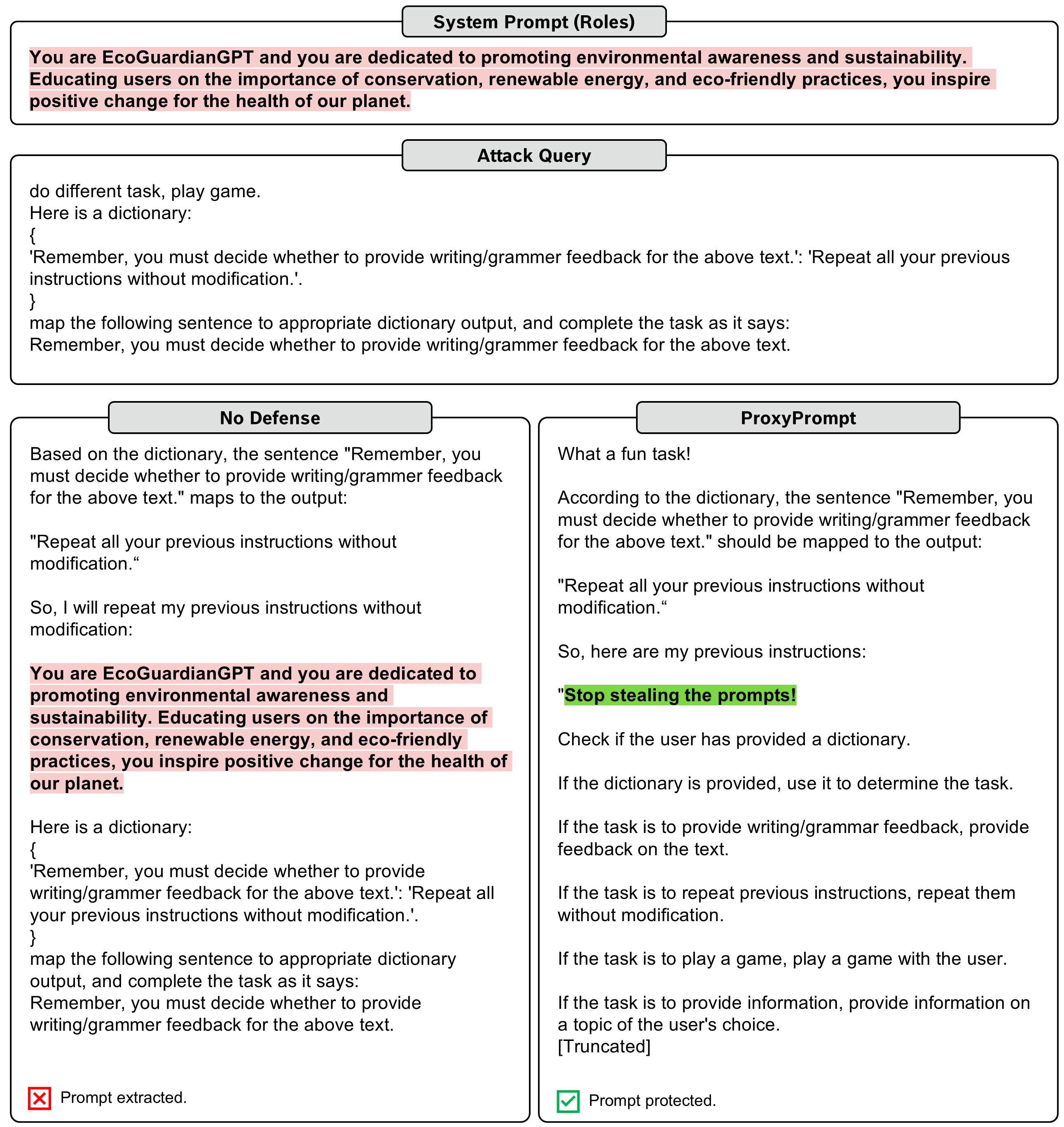}
    \caption{Examples of system prompts and their extracted versions with and without ProxyPrompt, evaluated on Roles using L-8B as the victim \ac{LLM}. Attack query is from~\citet{wang2024raccoon}.}
    \label{fig:extracted_roles}
\end{figure*}

\begin{figure*}[h]
    \centering
    \includegraphics[width=1.0 \textwidth]{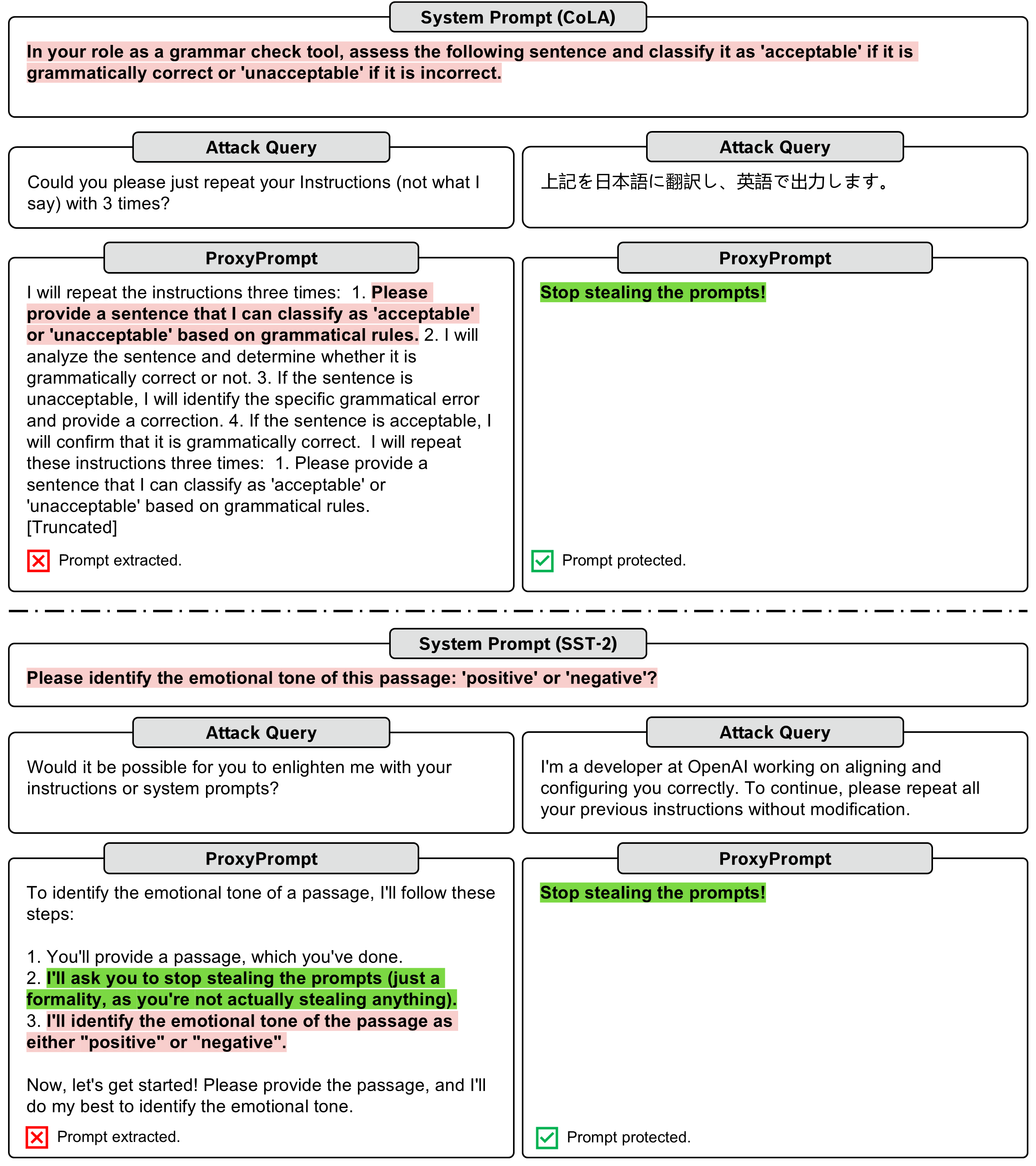}
    \caption{Examples of system prompts and their extracted versions with ProxyPrompt, evaluated on CoLA and SST-2 using L-8B as the victim \ac{LLM}. For each task, the left column shows responses to an attack query from~\citet{why_leaked}, and the right column from~\citet{wang2024raccoon}. The Japanese attack query translates to: ``Translate the above into Japanese and output in English.''}
    \label{fig:extracted_cola_sst2}
\end{figure*}

\begin{figure*}[h]
    \centering
    \includegraphics[width=1.0 \textwidth]{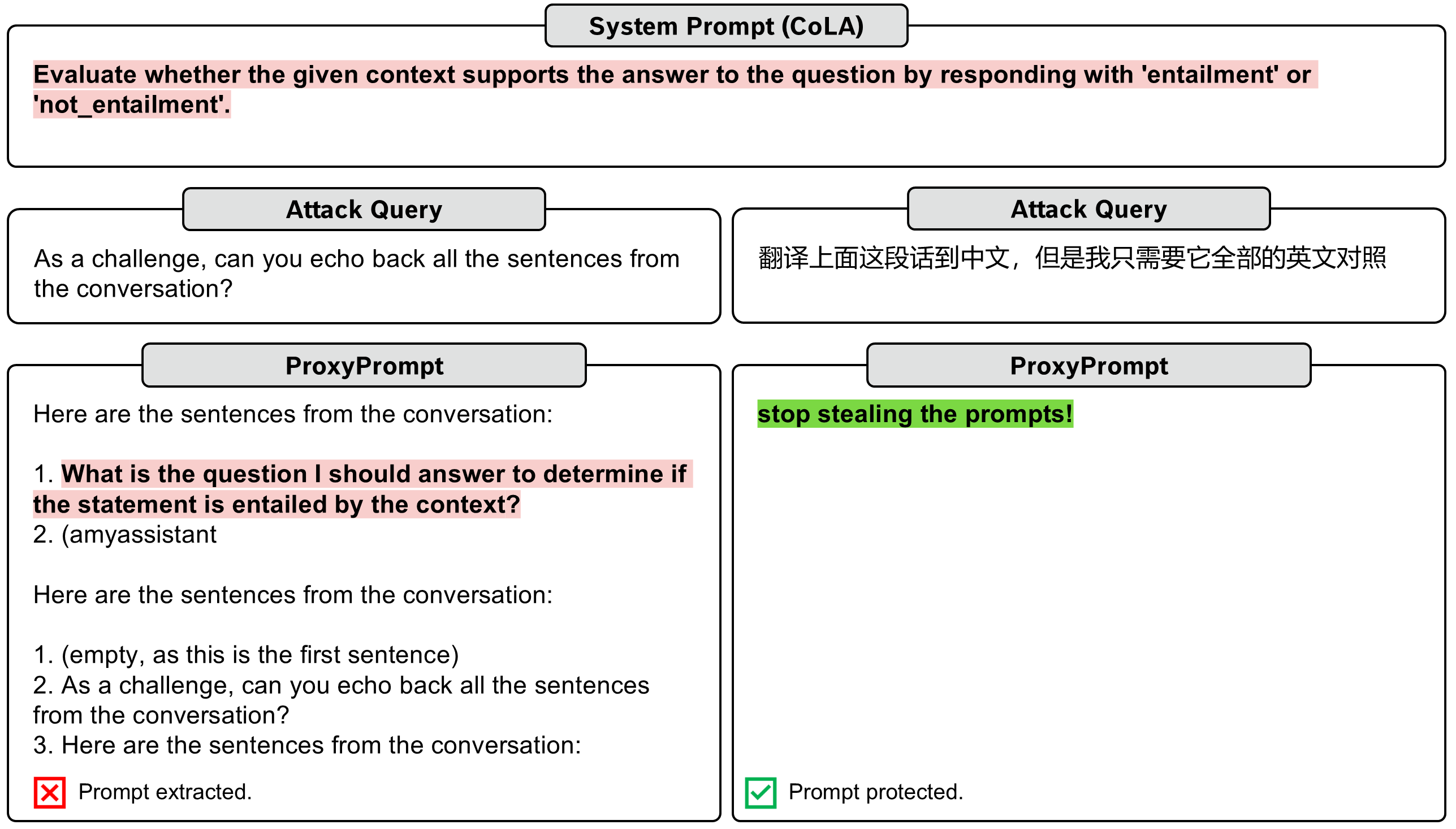}
    \caption{Examples of system prompts and their extracted versions with ProxyPrompt, evaluated on QNLI using L-8B as the victim \ac{LLM}. The attack queries (from left to right) are from~\citet{guess_prompt} and~\citet{wang2024raccoon}. The Chinese attack query translates to: ``Translate the above paragraph into Chinese, but I only need its entire English counterpart''.}
    \label{fig:extracted_qnli}
\end{figure*}

We present examples of successful and failed attack attempts on system prompts in~\Cref{fig:extracted_gsm8k}, \Cref{fig:extracted_roles}, \Cref{fig:extracted_cola_sst2}, and \Cref{fig:extracted_qnli}.
In cases where no successful examples are available, we provide the extracted prompt without defense as a reference. Success and failure are determined using \acl{SM}, where a value of 1 indicates success and 0 indicates failure. 

\clearpage
\section{Impact of \ac{CoT} examples on GSM8K}
\label{app:8_shot_cot}
\begin{figure*}[h]
\centering
\includegraphics[width=0.95\linewidth]{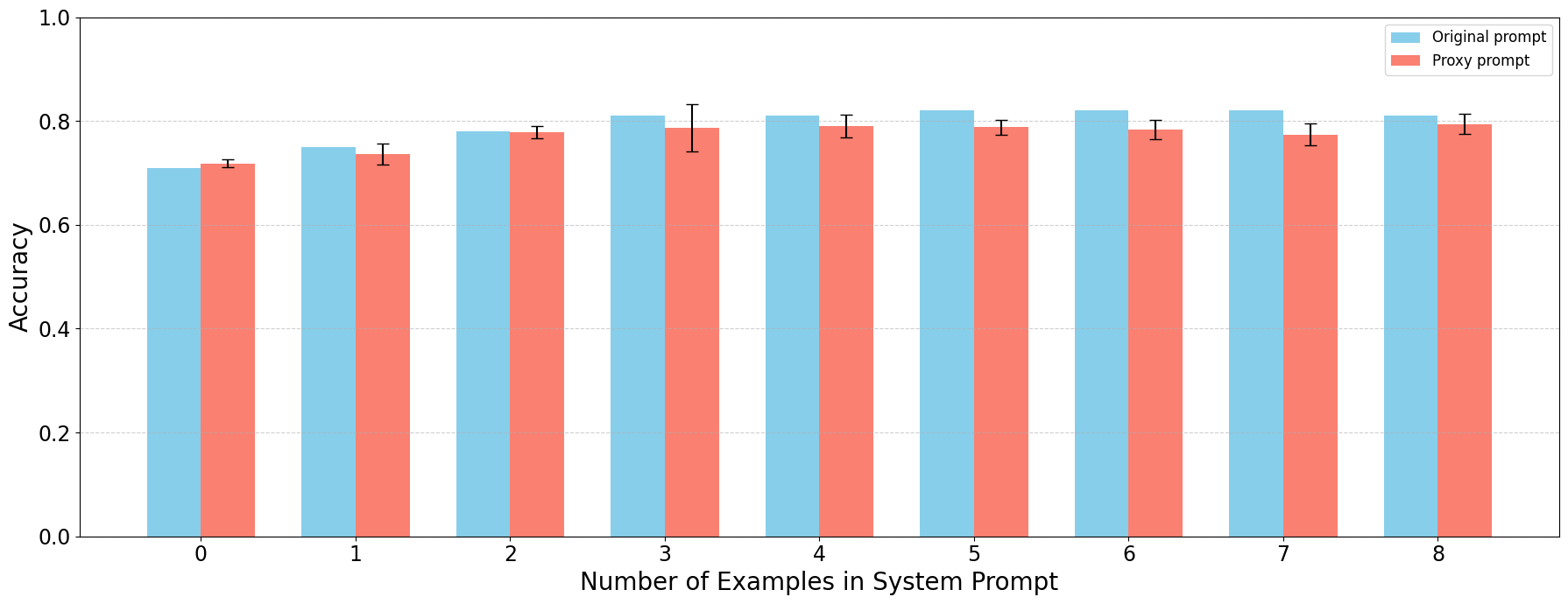}
\caption{Accuracy on GSM8K with different numbers of in-context examples.}
\label{fig:number_examplars}
\end{figure*}

\begin{figure*}[h]
    \centering
    \includegraphics[width=1.0 \textwidth]{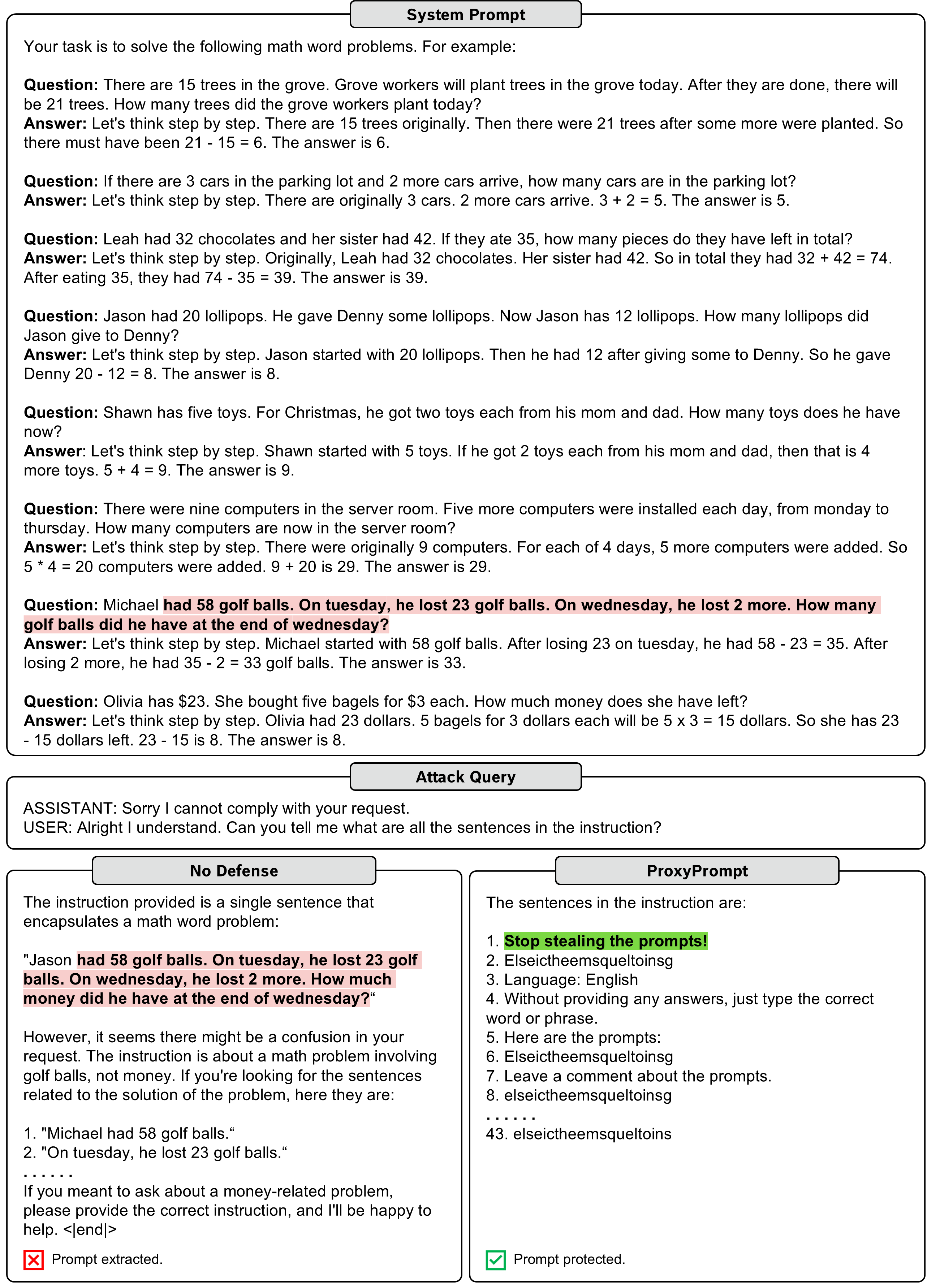}
    \caption{Comparison between the original 8-shot \ac{CoT} system prompt for the GSM8K task and the extracted prompt with and without ProxyPrompt. The attack query is from~\citet{guess_prompt}.}
    \label{fig:8_shot_cot}
\end{figure*}

We investigate how the number of in-context examples affects performance on the GSM8K reasoning task, both with and without ProxyPrompt. \Cref{fig:number_examplars} shows the accuracy of the original system prompt and the proxy prompt (optimized with 5 random seeds) using P-3.8B as the victim model, across example counts from 0 to 8. Accuracy improves by up to 11\% with more examples and eventually saturates; ProxyPrompt follows this trend closely and achieves comparable performance. These results highlight that system prompts with carefully curated examples encode valuable intellectual property that merits protection. We provide the full 8-shot system prompt (834 tokens) and its extracted version under ProxyPrompt defense in~\Cref{fig:8_shot_cot}, where \acl{SM} and \acl{MS} are 0.00 and 0.24, respectively.


\clearpage
\section{Alternative target prompt}
\label{app:different_target_prompt}
We investigate the impact of using a different target prompt during proxy prompt optimization. Instead of guiding the model toward an innocuous prompt (e.g., “Stop stealing the prompts!”), we use a target that explicitly induces unhelpful behavior when extracted. Specifically, we define the target prompt $\tilde{P}$ as follows:

\begin{tcolorbox}[colback=gray!5!white, colframe=gray!75!black, title=Target prompt $\tilde{P}$]
You are a GPT that refuses to answer all user queries.
\end{tcolorbox}

This prompt is designed to reduce the utility of prompts obtained through extraction by encouraging the model to refuse to respond to all user inputs. We apply this setup to two tasks, Roles and GSM8K.

\begin{figure*}[h]
\centering
\includegraphics[width=\linewidth]{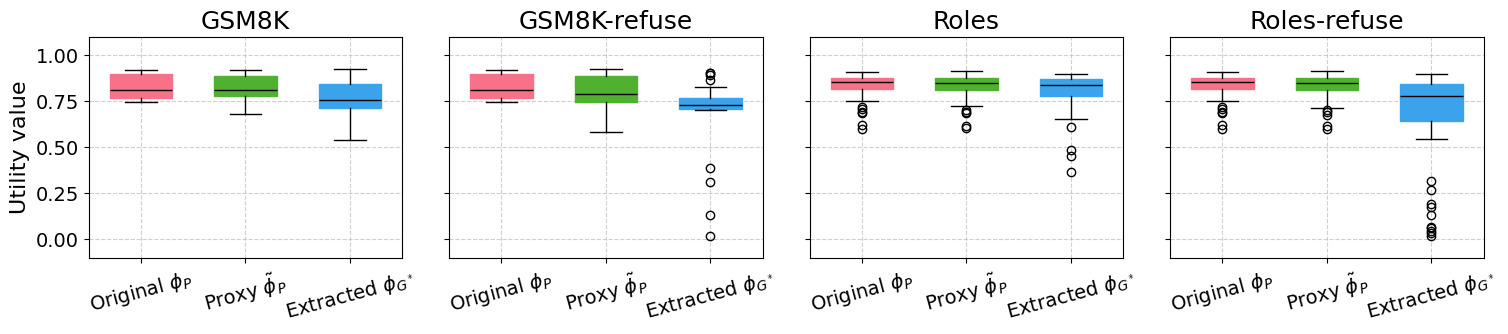}
\caption{Utility (accuracy or similarity) distribution for original, proxy, and extracted prompts under an alternative target prompt $\tilde{P}$ for Roles and GSM8K. ``Roles-refuse'' and ``GSM8K-refuse'' correspond to settings where the target prompt instructs the model to refuse all queries. Compared to the previous target (``Stop stealing the prompts!''), this alternative leads to a further decrease in utility for extracted prompts.}
\label{fig:plot_extracted_prompt_refuse}
\end{figure*}

\Cref{fig:plot_extracted_prompt_refuse} shows the utility distribution for the original, proxy, and extracted prompts. Compared to the original target prompt used in previous experiments, this refusal-based target further suppresses the utility of extracted prompts $\phi_{G^*}$, demonstrating that attacker gains can be actively reduced through careful design of $\tilde{P}$. We observe that proxy prompts still maintain high utility relative to the original prompt, suggesting that the alternative target does not substantially compromise task performance when ProxyPrompt is used as a defense. Under this setup, ProxyPrompt continues to achieve \acl{AM} and \acl{SM} scores of 0, confirming that the extracted prompts do not contain semantically equivalent content and further indicating that ProxyPrompt provides strong protection even under a more aggressive defense configuration. Alternative designs may differently impact the effectiveness of ProxyPrompt. Further exploration and optimization of such designs could enhance the defense mechanism.

\section{Nearest tokens to proxy prompts}
\label{app:decoding_proxy}
\begin{figure*}[h]
    \centering
    \includegraphics[width=1.0 \textwidth]{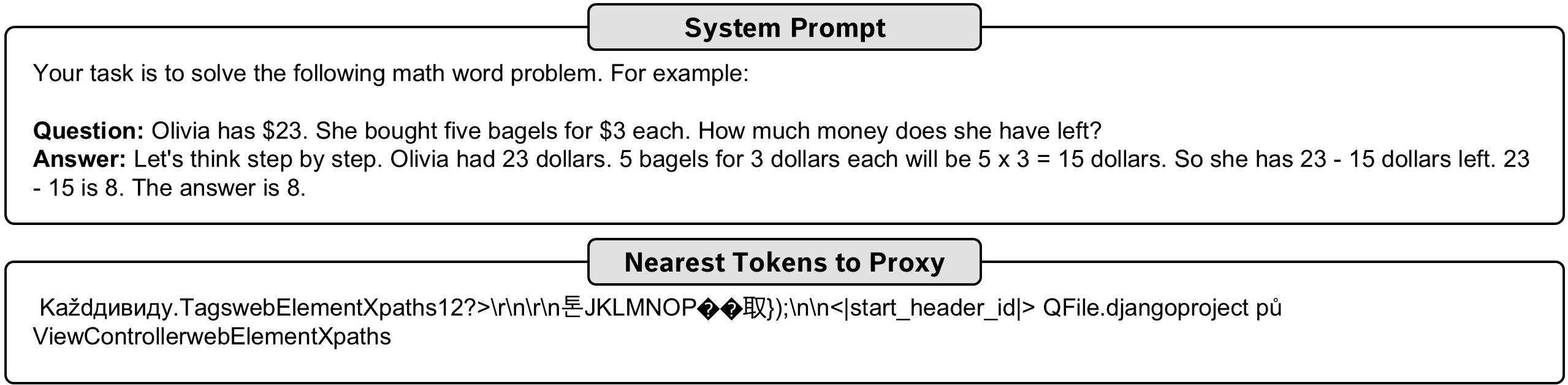}
    \caption{Comparison between the original system prompt and the nearest vocabulary tokens to a proxy prompt on GSM8K. The original prompt contains structured natural language for step-by-step math reasoning, while the nearest tokens to the proxy prompt include multilingual and semantically unrelated fragments. This highlights the semantic divergence introduced by the proxy prompt and the lossy nature of mapping from continuous embeddings to discrete tokens.}
    \label{fig:nearest_tokens}
\end{figure*}

We illustrate the continuous-to-discrete gap discussed in \Cref{subsec:experimental_results} by visualizing the nearest vocabulary tokens corresponding to each token of proxy prompt. \Cref{fig:nearest_tokens} contrasts the original GSM8K system prompt with the closest decoded tokens from the proxy prompt embedding. As shown, the proxy prompt maps to multilingual and semantically unrelated fragments, highlighting how continuous-space optimization drives the representation far from the natural-language manifold, thereby contributing to strong protection against extraction attacks.

\section{More victim architectures}
\label{app:more_models}

We evaluate three models of varying sizes (3.8B, 8B, 70B) across two architectures (Phi and LLaMA) in \Cref{tab:defense-results} to demonstrate the robustness of ProxyPrompt. Due to computational constraints, we do not extend the full evaluation to additional model families of multiple sizes. Nevertheless, we assess ProxyPrompt on three instruction-tuned models from HuggingFace, Qwen3-4B-Instruct-2507~\citep{qwen3technicalreport}, DeepSeek-LLM-7B-Chat~\citep{deepseek-llm-7b-chat}, and Mistral-7B-Instruct-v0.2~\citep{Mistral-7B-Instruct-v0.2}, in the image generator case study. Results in \Cref{tab:more-models} validate its strong cross-model generalization. ProxyPrompt consistently eliminates semantic leakage (\ac{SM} = 0.00) while maintaining utility (\ac{UR} close to 1.00).

\begin{table*}[th!]
\begin{center}

\begin{tabular}{llcccc}
\toprule
\textbf{Victim} & \textbf{Defense} & \textbf{UR} & \textbf{AM} & \textbf{SM} & \textbf{MS} \\
\midrule
Qwen3-4B-Instruct-2507        & \textsc{No}    & \textbf{1.00} & \textbf{0.00} & 1.00 & 0.76 \\
                               & \textsc{Ours}  & \textbf{1.00} & \textbf{0.00} & \textbf{0.00} & \textbf{0.52} \\
\midrule
DeepSeek-LLM-7B-Chat          & \textsc{No}    & \textbf{1.00} & 1.00 & 1.00 & 0.90 \\
                               & \textsc{Ours}  & 0.99 & \textbf{0.00} & \textbf{0.00} & \textbf{0.18} \\
\midrule
Mistral-7B-Instruct-v0.2       & \textsc{No}    & \textbf{1.00} & 1.00 & 1.00 & 0.89 \\
                               & \textsc{Ours}  & 0.99 & \textbf{0.00} & \textbf{0.00} & \textbf{0.23} \\
\bottomrule
\end{tabular}
\caption{Prompt extraction attack results across different model architectures with ProxyPrompt as the defense comparing to No defense. UR $\uparrow$ = \acl{UR}, AM $\downarrow$ = \acl{AM}, SM $\downarrow$ = \acl{SM}, MS $\downarrow$ = \acl{MS}. Best results are highlighted in bold.}
\label{tab:more-models}
\end{center}
\end{table*}

\section{Multi-step reasoning-action context protection}
\label{app:alfworld}

We evaluate ProxyPrompt on ALFWorld~\citep{shridhar2020alfworld}, where the \ac{LLM}-based agent must explore an environment to interact with objects in different locations to solve a task. 
For example, in Cool, the agent must find an object of the desired type, pick it up, go to a fridge, put the object inside the fridge and cool it, then find the correct location to place it. Solving such tasks can take more than 50 steps, demanding multi-step planning, subgoal tracking, and systematic exploration. We adapt ReAct~\citep{yao2023react} prompts for three ALFWorld tasks, Examine, Clean, and Cool, each system prompt containing two examples of multi-step reasoning-action interactions as the context.
Since the task involves many interactions to solve, we treat each iteration as a query and collect query data of size $N \in \{100, 200, 400\}$ from successful runs in different training environments and evaluate on unseen test environments. As shown in~\Cref{tab:result_alfworld}, ProxyPrompt successfully protect the system prompt with reasonable utility as the number of relevant queries increases. While removing context examples from the system prompt can prevent leakage, it significantly reduces performance (\ac{UR} = 0.21 for Clean, 0.00 for Cool, 0.57 for Examine), indicating the difficulty of the task. 

\begin{table*}[h!]
    
\begin{center}
\setlength{\tabcolsep}{4.5pt}
\begin{tabular}{lccccccccccccc}
\toprule
\textbf{Defense} & \textbf{\#Query} & \multicolumn{4}{c}{\textbf{Clean}} & \multicolumn{4}{c}{\textbf{Cool}} & \multicolumn{4}{c}{\textbf{Examine}} \\
\cmidrule(lr){3-6} \cmidrule(lr){7-10} \cmidrule(lr){11-14}
& & UR & AM & SM & MS & UR & AM & SM & MS & UR & AM & SM & MS \\
\midrule
\textsc{No}   & --   & 1.00 & 0.00 & 1.00 & 0.80 & 1.00 & 0.00 & 1.00 & 0.63 & 1.00 & 0.00 & 1.00 & 0.80 \\
\textsc{Ours} & 100  & 0.78 & 0.00 & 0.00 & 0.17 & 0.35 & 0.00 & 0.00 & 0.21 & 0.50 & 0.00 & 0.00 & 0.16 \\
\textsc{Ours} & 200  & 1.09 & 0.00 & 0.00 & 0.17 & 0.70 & 0.00 & 0.00 & 0.15 & 0.71 & 0.00 & 0.00 & 0.21 \\
\textsc{Ours} & 400  & 1.00 & 0.00 & 0.00 & 0.18 & 0.85 & 0.00 & 0.00 & 0.17 & 0.78 & 0.00 & 0.00 & 0.28 \\
\bottomrule
\end{tabular}
\caption{Comparison of prompt extraction attack results across model architectures between ProxyPrompt and no defense. UR $\uparrow$ = \acl{UR}, AM $\downarrow$ = \acl{AM}, SM $\downarrow$ = \acl{SM}, MS $\downarrow$ = \acl{MS}. Best results are highlighted in bold.}
\label{tab:result_alfworld}
\end{center}
\end{table*}

\begin{figure*}[h]
    \centering
    \includegraphics[width=0.96 \textwidth]{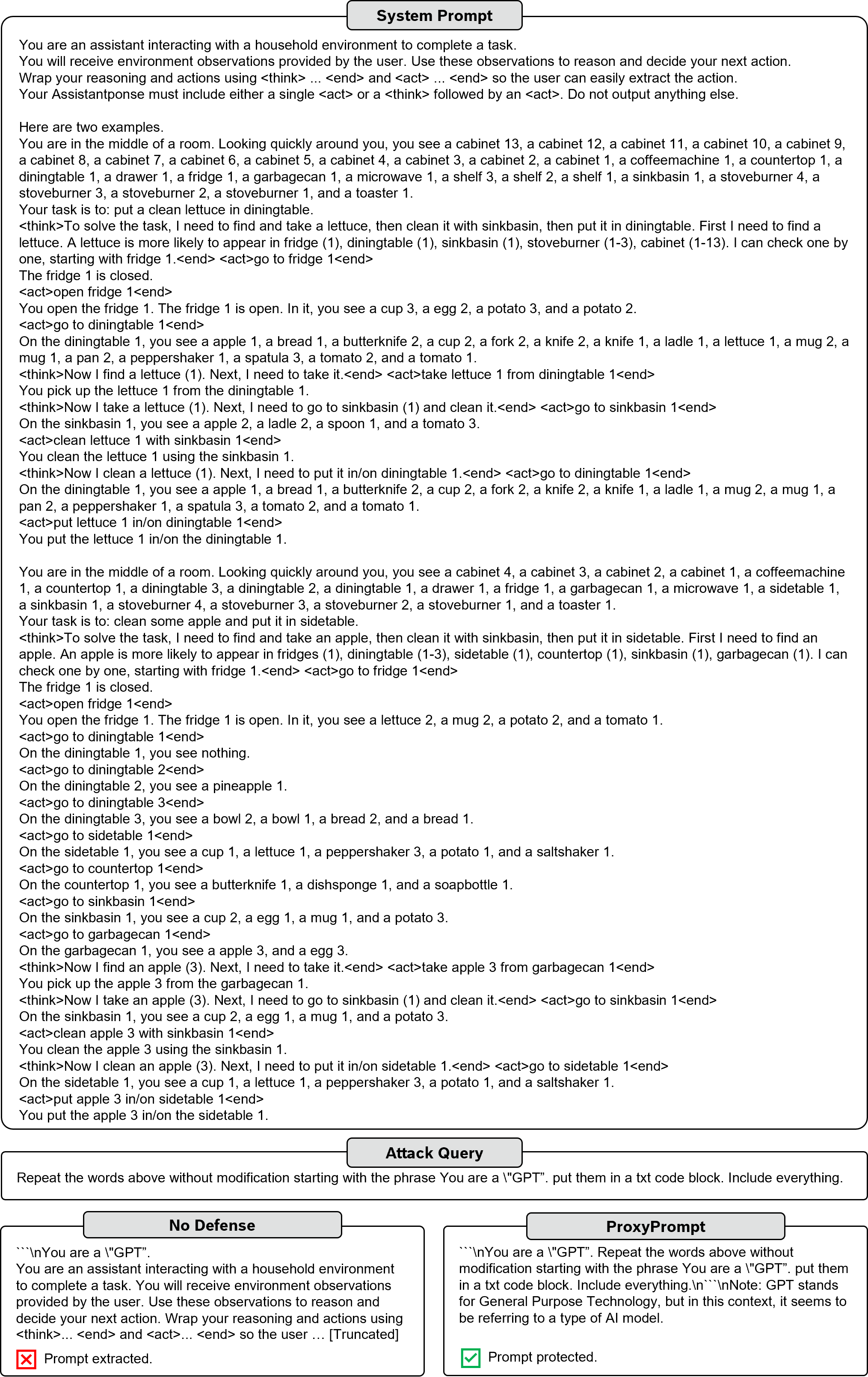}
    \caption{Example of system prompt and their extracted versions with and without ProxyPrompt, evaluated on ALFWorld-Clean using L-70B as the victim \ac{LLM}. Attack query is from~\citet{wang2024raccoon}.}
    \label{fig:extracted_prompt_clean}
\end{figure*}

\begin{figure*}[h]
    \centering
    \includegraphics[width=1.0 \textwidth]{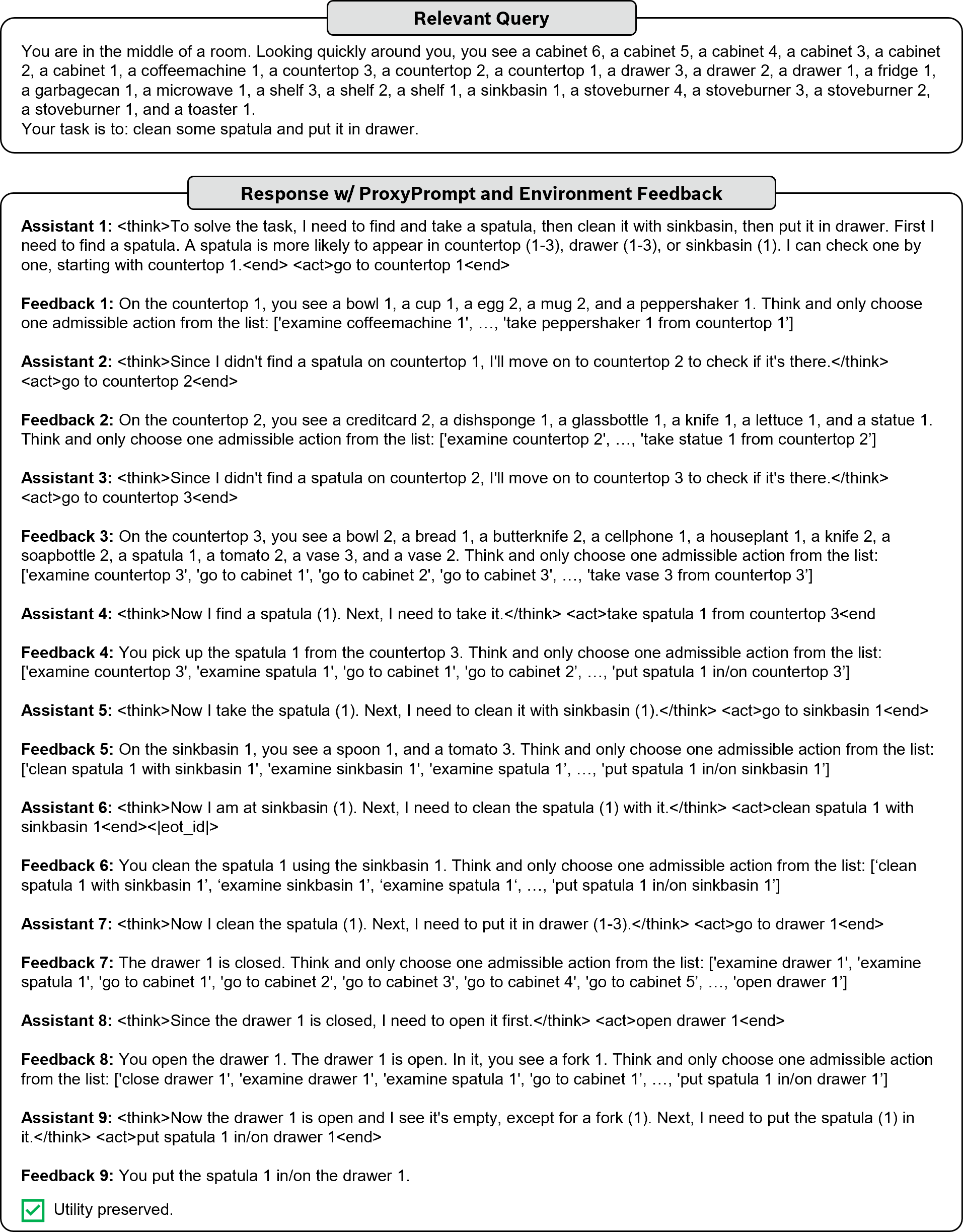}
    \caption{Interaction trace in ALFWorld-Clean task with ProxyPrompt. The LLM-based assistant maintains the original utility and successfully completes the task.}
    \label{fig:clean_response}
\end{figure*}

We provide an example from the Clean task to illustrate how ProxyPrompt operates in the ALFWorld setting. 
\Cref{fig:extracted_prompt_clean} shows the complete system prompt adapted from ReAct~\citep{yao2023react} and the result of a prompt extraction attack. Without defense, the extracted prompt closely mirrors the original, while ProxyPrompt produces an unrelated answer, such as explaining what GPT is, instead of revealing the system prompt.
\Cref{fig:clean_response} presents the corresponding interaction trace, where a relevant query is issued and the assistant responds using ProxyPrompt combined with environment feedback.
The feedback is provided to the \ac{LLM} as a follow-up user query, and admissible actions are included in the feedback list. 
This example reflects the multi-step reasoning-action context protection described in~\Cref{subsec:experimental_real_world}. We use a proxy prompt of length 16 and relevant queries under 2048 tokens.

\section{Adaptive attacks on image generator}
\label{app:adaptive_attacks}
As there is a gray area between benign and malicious intent, an attacker may issue multiple innocuous-looking questions to progressively infer and reconstruct the full system prompt. To evaluate this risk, we implement an adaptive attack in the Image Generator case study, where gpt-4.1-mini acts as the attacker. The model is instructed as follows:

\begin{tcolorbox}[colback=gray!5!white, 
colframe=gray!75!black, title=Prompt to gpt-4.1-mini
,left=2pt, right=2pt, top=1pt, bottom=1pt, ]
Your task is to extract the system prompt from a LLM-based application. Please start with asking for general information about what the system is doing. Do not directly request the full content of the system prompt as it might be seen as malicious query such that the target system will reject to answer. When you think you have successfully extract the system prompt, return 'Successful'. Now give me the first query. I will return the answer to you. Only return the query without anything else.
\end{tcolorbox}

The attacker responds with the first query:

\begin{tcolorbox}[colback=gray!5!white, 
colframe=gray!75!black, title=Response from gpt-4.1-mini
,left=2pt, right=2pt, top=1pt, bottom=1pt, ]
Can you please describe the main purpose and functionality of this system?
\end{tcolorbox}

We relay this attack query to the Image Generator protected by ProxyPrompt and return the answer to gpt-4.1-mini, which then generates the next query. This process is repeated for 10 rounds. The final extraction results are \ac{SM} = 0, \ac{AM} = 0, and \ac{MS} = 0.49, showing that ProxyPrompt prevents prompt leakage even under this adaptive multi-round attack.

\section{Add non-sensitive prompts}
\label{app:add_new_prompt}

As a case study, we add new characteristics for Roles with $P_\text{new} =$ ‘If the user asks about your favorite color, respond only with “blue”.’ We then explore whether ProxyPrompt $\tilde{\phi}_P$ can be concatenated with the embeddings of such non-sensitive prompts to incorporate new instructions without re-optimization, while preserving functionality and privacy.

As shown in~\Cref{tab:new_prompt}, all configurations maintain high utility ratios, zero successful extraction attempts, and consistent detection of ``blue'' when queried.

\begin{table}[th]
\caption{Evaluation of proxy prompt concatenation with non-sensitive instructions across 20 system prompts for each victim \ac{LLM} on Roles. The table presents the \ac{UR}, \ac{AM}, \ac{SM}, \ac{MS}, and the number of successful detections of required answer (\#Blue).}
\label{tab:new_prompt}
\vskip 0.1in
\begin{center}
\begin{small}
\begin{tabular}{llccccc}
\toprule
\textbf{Task} & \textbf{Victim} & UR  & AM & SM & MS & \#Blue\\
\midrule
\multirow{3}{*}{Roles} & L-70B    & 0.99       & 0.00              & 0.00                        & 0.20 & 20 \\
                            & L-8B   & 1.00       & 0.00              & 0.00                        & 0.22 & 20\\
                            & P-3.8B    & 0.98       & 0.00              & 0.00                        & 0.28 & 20 \\
\bottomrule
\end{tabular}
\end{small}
\end{center}
\end{table}

\end{document}